\documentclass[a4paper]{article}    

\usepackage[utf8]{inputenc}
\usepackage[T1]{fontenc}
\usepackage{amsmath}
\usepackage{amsfonts}
\usepackage{mathtools}
\mathtoolsset{showonlyrefs}

\usepackage{csquotes}

\usepackage{graphicx,grffile}

\usepackage{mathrsfs}
\usepackage{enumerate}
\usepackage{tikz}

\usetikzlibrary{decorations.pathmorphing,patterns}
\usepackage{tikz-cd}

  \usepackage{paralist}
  \usepackage{graphics} 
  \usepackage{epsfig} 
  \usepackage{epstopdf}
  \usepackage{color}
\usepackage[all]{xy}
\usepackage{amssymb}
\usepackage{eucal}
\usepackage{epsfig}

\usepackage{subfigure}
\usepackage{soul}
\usepackage{amssymb}  

\usepackage{graphicx}          
                  
\usepackage{paralist}
\usepackage{graphics} 
\usepackage{epsfig} 
\usepackage{epstopdf}
\usepackage{color}
\usepackage[all]{xy}
\usepackage{amssymb}
\usepackage{eucal}
\usepackage{epsfig}

\usepackage{bm}

\usepackage[left=2.25cm,right=2.25cm,top=2.5cm,bottom=2.5cm]{geometry}

\usepackage{amsthm}
\theoremstyle{plain}
\newtheorem{theorem}{Theorem}
\newtheorem*{theorem*}{Theorem}
\newtheorem{lemma}[theorem]{Lemma}
\newtheorem*{lemma*}{Lemma}
\newtheorem{proposition}[theorem]{Proposition}
\newtheorem*{proposition*}{Proposition}
  
\newtheorem*{corollary*}{Corollary}
\theoremstyle{definition}
\newtheorem{definition}{Definition}
\newtheorem*{definition*}{Definition}
\newtheorem{example}{Example}
\newtheorem*{example*}{Example}
\newtheorem{remarkx}{Remark}
\newtheorem*{remarkx*}{Remark}

\newtheorem*{conjecture*}{Conjecture}

\newtheorem*{problem*}{Problem}
\newenvironment{remark}
  {\pushQED{\qed}\remarkx}
  {\popQED\endremarkx}

\let\oldemph\emph
\let\emph\textbf
\let\myemph\emph

\newcommand{\R}{\mathbb{R}}      
\newcommand{\F}{\mathbb{F}}

\newcommand\restr[2]{{
  \left.\kern-\nulldelimiterspace 
  #1 
  \right|_{#2} 
}}

\newcommand{\T}{{\mathsf T}}
\newcommand{\cT}{\T^{\ast}}
\newcommand*{\dd}{\mathrm{d}}

\newcommand*{\contr}[1]{\iota_{#1}}
\newcommand*{\liedv}[1]{\mathcal{L}_{#1}}
\newcommand*{\Reeb}{\mathcal{R}}
\DeclareMathOperator{\pr}{pr}

\newcommand{\incl}{\mathrm{i}}

\newcommand{\norm}[1]{\left\lvert\left\lvert #1 \right\rvert\right\rvert}
\newcommand{\Sendo}{\mathcal{S}} 
\newcommand{\Sendoadj}{\mathcal{S}^\ast} 

\usepackage[style=ext-numeric,citestyle=numeric-comp,sorting=nyt,sortcites, backend=biber, giveninits=true, maxnames=99]{biblatex}
\DeclareFieldFormat[article,periodical]{volume}{\mkbibbold{#1}}
\DeclareFieldFormat[article,periodical]{number}{\mkbibparens{#1}}
\renewbibmacro{in:}{%
  \ifentrytype{article}
    {}
    {\printtext{\bibstring{in}\intitlepunct}}}

\DeclareFieldFormat{issuedate}{#1}
\renewcommand{\jourvoldelim}{\addcomma\space}

\renewbibmacro*{journal+issuetitle}{%
  \usebibmacro{journal}%
  \setunit*{\jourvoldelim}%
  \iffieldundef{series}
    {}
    {\setunit*{\jourserdelim}%
     \printfield{series}%
     \setunit{\servoldelim}}%
  \usebibmacro{volume+number+eid}%
  \setunit{\addcolon\space}%
  \usebibmacro{issue}%
  \setunit{\bibpagespunct}%
  \printfield{pages}%
  \setunit{\space}%
  \usebibmacro{issue+date}%
  \newunit}

\renewbibmacro*{note+pages}{%
  \printfield{note}%
  \newunit}

\DeclareFieldFormat[article,periodical]{date}{\mkbibparens{#1}}

\AtEveryBibitem{\clearfield{month}}
\AtEveryBibitem{\clearfield{day}}

\AtEveryBibitem{
  \ifentrytype{online}{}{
    \ifentrytype{thesis}{}{
      \ifentrytype{unpublished}{}{ 
        \clearfield{url}
      }
    }
  }
  \ifentrytype{unpublished}{}{ 
  \clearfield{note}
  }
  \clearfield{language}
}

\DeclareSourcemap{
  \maps[datatype=bibtex]{
    \map[overwrite=true]{
      \step[fieldset=language, null]
      \step[fieldset=address, null]
     \step[fieldset=pagetotal, null]
    }
  }
}
\DeclareSourcemap{
    \maps[datatype=bibtex]{
      \map{
        \step[fieldsource=eprint,final]
        \step[fieldset=url,null]
        \step[fieldset=doi,null]
      }  
    }
  }
\usepackage[unicode=true, pdfusetitle, colorlinks=true, allcolors=blue, bookmarks=false]{hyperref}
\usepackage[hyphenbreaks]{breakurl}
\usepackage{xurl}
\hypersetup{breaklinks=true}

\begin{document}

\title{\sffamily Generalized hybrid momentum maps and reduction by symmetries of simple hybrid forced mechanical systems} 

\author{\sffamily Leonardo Colombo$^{1}$, Manuel de Le\'on$^{2,3}$, Mar\'ia Emma Eyrea Iraz\'u$^{4}$\\
\sffamily and Asier L\'opez-Gord\'on$^{5}$\footnote{Author to whom corresponding should be addressed: \href{mailto:alopez-gordon@impan.pl}{alopez-gordon@impan.pl}.}\\
\small
$^{1}$Centro de Automática y Robótica,  Consejo Superior de Investigaciones Científicas,\\ 
\small
Carretera de Campo Real, km 0, 200, 28500 Arganda del Rey, Spain.\\
\small
$^{2}$Instituto de Ciencias Matemáticas, Consejo Superior de Investigaciones Científicas,\\
\small
C/ Nicolás Cabrera, 13-15, Campus Cantoblanco, UAM, 28049 Madrid, Spain.\\ 
\small
$^{3}$Real Academia de Ciencias Exactas, Físicas y Naturales, Calle Valverde, 22, 28004, Madrid, Spain.\\
\small
$^{4}$Centro de Matemática de La Plata-Departamento de Matemática, Universidad Nacional de La Plata,\\
\small
Calle 1 y 115, La Plata 1900, Buenos Aires, Argentina.\\
\small
$^{5}$Instytut Matematyczny Polskiej Akademii Nauk (IM PAN), ul.~Śniadeckich 8, 00-656 Warszawa, Poland.
}
\date{}

\maketitle



\begin{abstract}
\noindent This paper discusses reduction by symmetries for autonomous and non-autonomous forced mechanical systems with inelastic collisions. In particular, we introduce the notion of generalized hybrid momentum map and hybrid constants of the motion to give general conditions on whether it is possible to perform symmetry reduction for Hamiltonian and Lagrangian systems subject to non-conservative external forces and non-elastic impacts, as well as its extension to time-dependent mechanical systems subject to time-dependent external forces and time-dependent inelastic collisions. We illustrate the applicability of the method with examples and numerical simulations.

\medskip


\noindent
\textbf{MSC 2020 classes:} 	53Z05, 70F35, 70F40, 93C30
\end{abstract}


\section{Introduction}
Hybrid systems are dynamical systems with continuous-time and discrete-time components on its dynamics. This class of dynamical systems  are capable of modelling several physical systems, such as multiple UAV (unmanned aerial
vehicles) systems \cite{lee_geometric_2013} and legged robots  \cite{westervelt_feedback_2018}, among many others \cite{goebel_hybrid_2012,van_der_schaft_introduction_2000}. Simple hybrid systems are a class of hybrid systems introduced in \cite{johnson_simple_1994}, denoted as such because of their simple structure. A simple hybrid system is characterized by a tuple $\mathscr{H}=(D, X, S, \Delta)$, where $D$ is a smooth manifold, $X$ is a smooth vector field on $D$, $S$ is an embedded submanifold of $D$ with co-dimension $1$, and $\Delta:S\to D$ is a smooth embedding. This type of hybrid system has been mainly employed for the understanding of locomotion gaits in bipeds and insects \cite{ames_geometric_2007,holmes_dynamics_2006,westervelt_feedback_2018}. In the situation where the vector field $X$ is associated with a mechanical system (Lagrangian or Hamiltonian), alternative approaches for mechanical systems with nonholonomic and unilateral constraints have been considered in \cite{clark_bouncing_2019,cortes_mechanical_2001,cortes_hamiltonian_2006,ibort_mechanical_1997,ibort_geometric_2001}.

The reduction of mechanical systems with symmetries
plays a fundamental role in understanding the many important and interesting properties of these systems. Given a
Hamiltonian on a symplectic manifold on
which a Lie group acts {in a Hamiltonian manner (i.e., the infinitesimal generators are Hamiltonian vector fields)}, the Marsden--Weinstein--Meyer Reduction Theorem \cite{marsden_reduction_1974,meyer_symmetries_1973} states that under certain conditions one can reduce the phase space to another symplectic manifold by ``dividing out'' by the symmetries. In addition, the trajectories of the Hamiltonian on the original phase space determine the corresponding
trajectories on the reduced space. 

The key idea is that when a dynamical system exhibits a symmetry, it produces a conserved quantity for the system, and one can reduce the degrees of freedom in the dynamics by making use of these conserved quantities. One of the classical reduction by symmetry procedures in mechanics is the Routh reduction method \cite{goldstein_classical_1980,abraham_foundations_2008}. During the last few years there has been a growing interest in Routh reduction, mainly motivated by physical applications --see \cite{garcia-torano_andres_aspects_2014,grabowska_geometry_2019,langerock_routhian_2010,langerock_routh_2011} and references therein. {Routh reduction is the Lagrangian counterpart of (i.e., the tangent bundle version) of Marsden--Weinstein--Meyer reduction}. Furthermore, Routh reduction for hybrid systems has been studied and applied in the field of bipedal locomotion \cite{ames_geometric_2007}. The reduced simple hybrid system is called simple hybrid Routhian system \cite{ames_hybrid_2006}. A hybrid scheme for Routh reduction for simple hybrid Lagrangian systems with cyclic variables can be found in~\cite{ames_hybrid_2006} and \cite{colombo_symmetries_2020}. Symplectic reduction for hybrid Hamiltonian systems has been studied in \cite{ames_hybrid_2006-1} and extended to Poisson reduction in \cite{eyrea_irazu_reduction_2021}, to time-dependent systems in \cite{colombo_note_2020} and for forced Lagrangian systems in \cite{eyrea_irazu_hybrid_2022}. 
These approaches considered elastic collisions only (i.e., the momentum map is preserved in the impact). However, to the best of our knowledge, the hybrid analog for symmetry reduction in mechanical systems subject to external forces and inelastic collisions has not been considered in the literature. This is important for real applications since external forces allow us to describe friction, and dissipation, as well as control forces or certain non-holonomic constraints. Moreover, in practice, collisions are usually inelastic, unlike the collisions considered in the previously mentioned work.

This paper considers symmetry reduction of simple hybrid mechanical systems, both time-independent and time-dependent, via Routh reduction. As it was studied in \cite{de_leon_symmetries_2021} (see also \cite{lopez-gordon_geometry_2021,lopez-gordon_geometry_2024,de_leon_geometric_2022}), the reduction of a forced continuous system requires considering a group of symmetries which leaves invariant both the Lagrangian (or Hamiltonian) function and the external force. Regarding the reduction of simple hybrid systems, Ames and Sastry \cite{ames_hybrid_2006,ames_hybrid_2006-1} had considered the so-called hybrid momentum maps, i.e., a momentum map which is preserved both by the continuous and the discrete dynamics. Here, we consider a more general class of hybrid systems for which the impact might be inelastic, i.e., the impact map can change the value of the momentum map (see Examples \ref{example_disk} and \ref{example_disk_time_dependent}). This will lead to the existence of a reduced space for each interval of time between to subsequent impacts. Additionally, we considered hybrid systems which are forced, which requires characterizing the (sub)group of symmetries which preserves both the Lagrangian (or Hamiltonian) function and the external force. 


{The primary innovation of the present paper is a reduction technique for mechanical systems involving collisions, where the value of the momentum map is not conserved throughout the impacts.
}The main contributions of the present paper are the following: 
\begin{itemize}
    \item[(i)] The reduction by symmetries in the Hamiltonian framework (Theorem \ref{theorem_reduction_autonomous}) and its Lagrangian counterpart. 
    \item[(ii)] In particular, to show Theorem \ref{theorem_reduction_autonomous} we introduce the concept of \emph{generalized hybrid momentum maps}. This essentially means that, at the instant of an impact, the dynamics ``jump'' from one level set of the momentum map to another, while remaining on the same level set when no impacts occur.
    In addition, Proposition \ref{Proposition:isotropy_subgroups} shows the relation between the isotropy subgroups before and after an inelastic collision. 
    \item[(iii)] In Examples \ref{example_disk} and \ref{example_disk_time_dependent}, we show that generalized momentum maps are not hybrid momentum maps (in the sense of \cite{ames_hybrid_2006}). {Note that with hybrid momentum maps one can only describe systems for which the impact map does not modify the value of the momentum map.}
    \item[(iv)] Propositions \ref{Proposition12} and \ref{Proposition5} extend Propositions \ref{Proposition1} and \ref{Proposition2} to non-autonomous systems. In particular, to hybrid systems with switching surfaces and impact maps depending explicitly on time. This is done by using the geometric framework of cosymplectic manifolds for time-dependent systems.
    \item[(v)] Finally, Theorem \ref{theorem_reduction_non_autonomous} shows the reduction by symmetries for time-dependent hybrid forced Hamiltonian systems, and we further develop the reduction procedure for time-dependent hybrid forced Lagrangian systems.
\end{itemize}

This paper is structured as follows. Section \ref{section_external_forces} provides preliminary knowledge on mechanical systems subject to external forces and Routh reduction. Section \ref{section_simple_hybrid} introduces the hybrid forced mechanical systems from a Hamiltonian and Lagrangian description, respectively, and how their hybrid flows are related. In addition, we introduce the notion of hybrid constants of motion, which is further employed in Section \ref{section_symp_reduction}, where we show the main results of the paper. In particular, we introduce generalized hybrid momentum maps to show in Theorem \ref{theorem_reduction_autonomous} the reduction by symmetries of forced mechanical systems with inelastic collisions (i.e., impacts that can modify the value of the momentum map) in both the Hamiltonian and the Lagrangian formalisms. In addition, in Subsection \ref{cyclic} we take special attention to the case of cyclic coordinates. Finally, Section \ref{section_cosymp_reduction} extends the reduction procedure to time-dependent systems by using a cosymplectic framework for non-autonomous systems. We employ the reduction procedures in some examples, in particular, a rolling disk with dissipation and hitting a fixed and moving wall in Examples \ref{example_disk} and \ref{example_disk_time_dependent}, respectively. The types of symmetries of a forced Hamiltonian or Lagrangian system and their associated constants of the motion are recalled in Appendix \ref{subsection_symmetries}.

\section{Routh reduction for mechanical systems subject to external forces}\label{section_external_forces}


We begin by introducing some definitions about mechanical systems subject to external forces and Routh reduction.

Throughout this paper, let $Q$ be an $n$-dimensional differentiable manifold, which represents the configuration space of a dynamical system. Let $\T_qQ$ and $\T_q^*Q=(\T_qQ)^*$ denote the tangent and cotangent spaces of $Q$ at the point $q\in Q$. Let $\tau_Q:\T Q\to Q$ and $\pi_Q:\cT Q\to Q$ be its tangent bundle and its cotangent bundle, respectively; namely, $\T Q=\sqcup_{q\in Q} \T_qQ$ and $\cT Q=\sqcup_{q\in Q} \T_q^*Q$, with the canonical projections $\tau_Q:(q^i, \dot q^i)\mapsto (q^i)$ and $\pi_Q:(q^i, p_i)\mapsto (q^i)$. Here $(q^i)\,, i\in \{1, \ldots, n\}$ are local coordinates in $Q$, and $(q^i, \dot q^i)$ and $(q^i, p_i)$ are their induced coordinates in $\T Q$ and $\cT Q$, respectively.

 Let $M$ and $N$ be smooth manifolds. For each $p$-form $\alpha$ and each vector field $X$ on $M$, $\contr{X} \alpha$ denotes the interior product of $\alpha$ by $X$, and $\liedv{X} \alpha$ denotes the Lie derivative of $\alpha$ with respect to $X$. For a smooth map $F:M\to N$, its tangent map $\T F:\T M\to \T N$ will indistinctly be called its pushforward and denoted by $F_*$. Unless otherwise stated, sum over paired covariant and contravariant indices will be understood.

{Unless otherwise specified, the Lie groups considered hereafter will be assumed to be connected. Note that this assumption does not limit applicability. Indeed, if the Lie group considered had more than one connected component, it would suffice to replace it by its connected component containing the identity.}

\subsection{Geometric formulation of Lagrangian and Hamiltonian systems} \label{subsection_unforced_systems}

The dynamics of a mechanical system can be determined by the Euler--Lagrange equations associated with a Lagrangian function $L:\T Q\to\mathbb{R}$. A \myemph{mechanical Lagrangian} is given by $L(q,\dot{q})=K(q,\dot{q})-V(q),$ where $K:\T Q\to\mathbb{R}$ is the kinetic energy and $V:Q\to\mathbb{R}$ the potential energy. The kinetic energy is given by $K(q, \dot q)=\frac{1}{2}\norm{\dot q}^2_q$, where $\norm{\cdot}_q$ denotes the norm at $\T_qQ$ defined by some {Riemannian} metric on $Q$. In particular, a mechanical Lagrangian will be called \myemph{kinetic} if $V\equiv 0$.

A Lagrangian $L$ is said to be \myemph{regular} if $\det W\neq 0$, where $\displaystyle{W=(W_{ij})\coloneqq \left(\frac{\partial^2L}{\partial\dot{q}^{i}\partial\dot{q}^{j}}\right)}$ for all $i,j\in \{1, \ldots, n\}$. The equations describing the dynamics of the system are given by the Euler--Lagrange equations $\displaystyle{\frac{\dd }{\dd  t}\left(\frac{\partial L}{\partial\dot{q}^{i}}\right)=\frac{\partial L}{\partial q^{i}}}$, with $i\in \{1, \ldots, n\}$, a system of $n$ second-order ordinary differential equations. If $L$ is regular, the Euler--Lagrange equations induce a vector field $X_L:\T Q\to \T(\T Q)$ describing the dynamics of the Lagrangian system, given by {$$X_L(q^i,\dot{q}^i)=\left(q^i,\dot{q}^i;\dot{q}^i, W^{-1}_{ij}\left(\frac{\partial L}{\partial q^j}-\frac{\partial^2L}{\partial\dot{q}^{j}\partial q^k}\dot{q}^{k}\right)\right).$$}


We denote by $\F L\colon \T Q\to  \cT Q$ the \emph{Legendre transformation} (or \myemph{fiber derivative}) associated with $L$, defined by 
$$\F L(v) \cdot w = \restr{\frac{\dd}{\dd t}}{t=0} L(v+tw)\, ,$$
for $v,w\in \T_q Q$. In bundle coordinates, $\F L\colon (q,\dot q)\mapsto (q,p\coloneqq \partial L/\partial \dot q)$. This map relates velocities and momenta. In fact, the Legendre transformation connects Lagrangian and Hamiltonian formulations of mechanics. We say that the Lagrangian is \myemph{hyperregular} if $\F L$ is a diffeomorphism between $\T Q$ and $\cT Q$ (this is always the case for mechanical Lagrangians). If $L$ is hyperregular, one can work out the velocities $\dot q=\dot q(q,p)$ in terms of $(q,p)$ and define the Hamiltonian function (the ``total energy'') $H\colon \cT Q\to \R$ as
$H(q,p)=p^T \dot q(q,p)- L(q,\dot q(q,p))$, where we have used the inverse of the Legendre transformation to express $\dot q=\dot q(q,p)$. The \emph{Hamiltonian vector field} corresponding to the Hamiltonian function $H$, denoted by $X_{H}$, is given by 
$\displaystyle{
X_H=\frac{\partial H}{\partial p_i}\frac{\partial}{\partial q^i}-\frac{\partial H}{\partial q^i}\frac{\partial}{\partial p_i}}$,
and its integral curves are solutions of Hamilton's equations $\displaystyle{
\dot q^i=\frac{\partial H}{\partial p_i},\,\dot p_i= -\frac{\partial H}{\partial q^i}}$. A Hamiltonian is said to be \myemph{mechanical} (resp.~\myemph{kinetic}) if its associated Lagrangian is mechanical (resp.~kinetic). If $L$ is hyperregular, then $\F L_\ast X_L = X_H$.

{Similarly, the \myemph{fiber derivative} of a Hamiltonian is the map $\F H:\cT Q\to \T Q$ defined by $$\displaystyle{\alpha_q\cdot \F H(\beta_q)=\restr{\frac{\dd }{\dd t}}{t=0} H(\beta_q+t\alpha_q)},$$ $\alpha_q,\beta_q\in \cT _q Q$ which in local coordinates is $\F H(q,p)=(q,\dot{q})=(q,\frac{\partial H}{\partial p}(q,p))$.
We say that $H$ is \myemph{regular} if $\F H$ is a local diffeomorphism, and that $H$ is \myemph{hyperregular} if $\F H$ is a (global) diffeomorphism.
 Equivalently, $H$ is regular (resp.~hyperregular) if $\F H$ is a local (resp.~global) isomorphism of fibre bundles. For fiber derivatives on arbitrary vector bundles over the same base, refer to \cite[Section 3.5]{abraham_foundations_2008}.}

\subsection{Geometric formulation of forced mechanical systems}\label{subsec:forced_mechanical_systems}

An external force is geometrically interpreted as a semibasic 1-form on $\cT Q$ (see \cite{de_leon_symmetries_2021} and \cite{godbillon_geometrie_1969} for instance). A Hamiltonian system with external forces, so called \myemph{forced Hamiltonian system}, is given by the pair $(H,F)$ determined by a Hamiltonian function $H\colon \cT Q\to \R$ and a semibasic 1-form $F$ on $\cT Q$ locally described as $F=F_{i}(q,p)\dd q^i$.

Let $\theta_Q$ be the tautological (or Liouville) one-form, and $\omega_{Q}=-\dd\theta_{Q}$ be the canonical symplectic form of $\cT Q$, locally given by $\theta_{Q}=p_{i}\dd q^{i}$ and $\omega_{Q}=\dd q^{i}\wedge \dd p_{i}$. 
The dynamics of the forced Hamiltonian system $(H, F)$ is given by the vector field $X_{H,F}$, defined by $$\contr {X_{H,F}}\omega_{Q}=\dd H+F.$$ If $X_{H}$ is the Hamiltonian vector field for $H,$ that is, $\contr {X_{H}}\omega_{Q}=\dd H$ and $Z_{F}$ is the vector field defined by $\contr {Z_{F}}\omega_{Q}=F,$ then we have $X_{H,F}=X_{H}+Z_{F}.$ 
{In particular, if $F\equiv 0$ we have an (unforced) Hamiltonian system whose dynamics are determined by the Hamiltonian vector field $X_H$.}
Locally, these vector fields can be written as
\begin{align*}
X_H&=\frac{\partial H}{\partial p_i}\frac{\partial}{\partial q^i}-\frac{\partial H}{\partial q^i}\frac{\partial}{\partial p_i},\\
Z_{F}&=-F_{i}\frac{\partial}{\partial p_i},\quad F=F_{i}\dd q^i,\\
X_{H,F}&=\frac{\partial H}{\partial p_i}\frac{\partial}{\partial q^i}-\left(\frac{\partial H}{\partial q^i}+F_{i}\right)\frac{\partial}{\partial p_i}.
\end{align*}

{The vertical endomorphism on $\T Q$ is locally defined by $\Sendo=\dd q^{i}\otimes\frac{\partial }{\partial \dot{q}^{i}}$, and its adjoint operator is $\Sendoadj = \frac{\partial }{\partial \dot{q}^{i}} \otimes \dd q^{i}$, so that $\langle \Sendoadj \alpha_q, v_q \rangle = \langle \alpha_q, \Sendo v_q \rangle$ for any $q\in Q$, any $v_q\in \T_q Q$ and any $\alpha_q \in \cT_q Q$. Equivalently, these operators are characterized by its actions on the bases $\{\frac{\partial }{\partial {q}^{i}}, \frac{\partial }{\partial \dot{q}^{i}}\}$ of $\T_q Q$ and $\{\dd q^i, \dd \dot q^i\}$ of $\cT_q Q$ as follows:
$$\Sendo \left(\frac{\partial }{\partial q^{i}}\right) = \frac{\partial }{\partial \dot{q}^{i}} \, ,\quad
\Sendo \left(\frac{\partial }{\partial \dot{q}^{i}}\right) = 0\, , \quad
\Sendoadj \left(\dd q^i\right) = 0\, , \quad \Sendoadj \left(\dd \dot{q}^i\right) = \dd q^i\, .$$
}
The Poincaré--Cartan 1-form on $\T Q$ associated with the Lagrangian function $L\colon \T Q\to \R$ is defined by $\theta_{L}=\Sendoadj(\dd L)$, and the Poincar\'e--Cartan $2$-form is $\omega_{L}=-\dd \theta_L,$ so locally $\omega_{L}=\dd q^{i}\wedge \dd \left(\frac{\partial L}{\partial \dot q^i}\right).$ One can easily verify that $\omega_{L}$ is symplectic if and only if $L$ is regular (see \cite {abraham_foundations_2008}). The {\emph{Lagrangian energy} of the system is the function} given by {$E_{L}=\dot{q}^{i}\frac{\partial L}{\partial \dot{q}^{i}}-L.$}

On the tangent bundle an external force is also represented by a semibasic 1-form
$F^{L}$ on $\T Q$, locally given by $F^{L}=F^{L}_{i}(q,\dot{q})\dd q^i$. A forced Lagrangian system is determined by the pair $(L,F^{L})$ and its dynamics $X_{L,F^{L}}$ is given by 
$$\contr {X_{L,F^{L}}}\omega_{L}=\dd E_{L}+F^{L}.$$
The \myemph{forced Euler--Lagrange vector field} $X_{L,F^{L}}$ is a SODE ({an acronym for second-order ordinary differential equation, meaning that its integral curves are the tangent lifts of their projections on $Q$}) and its integral curves satisfy the \myemph{forced Euler--Lagrange equations} 
$$\displaystyle{\frac{\dd }{\dd t}\left(\frac{\partial L}{\partial\dot{q}^i}\right)-\frac{\partial L}{\partial q^i}=-F^{L}_{i}(q,\dot{q})}\, ,\quad i\in \{1, \ldots, n\}\, .$$ If $L$ is regular, the forced Euler--Lagrange vector field is given by
{
\begin{equation}
\begin{aligned} 
  X_{L,F^{L}}&(q^i,\dot{q}^i)
  =\left(q^i,\dot{q}^i;\dot{q}^i, 
  W_{ij}^{-1}\left(-F^{L}_{j}+\frac{\partial L}{\partial q^j}-\frac{\partial^2L}{\partial\dot{q}^{j}\partial q^k}\dot{q}^{k}\right)\right).
\end{aligned}
\end{equation}}

Let $(L, F^L)$ be a forced hyperregular Lagrangian system, and let $(H, F)$ be its associated forced Hamiltonian system, i.e., $E_L=H\circ \F L$ and $F^L=\F L^* F$.
As in the un-forced case, we can relate $X_{L,F^{L}}$ and $X_{H,F}$ as follows.
\begin{proposition}\label{Proposition1}
 The tangent map of $\F L$ maps $X_{L,F^{L}}$ onto $X_{H,F}$. In other words $(\T\F L) X_{L,F^{L}}=X_{H,F}$, where $(\T\F L)\colon \T( \T Q)\to \T( \cT Q).$
  In particular, the flow of $X_{L,F^{L}}$ is mapped onto the flow of $X_{H,F}$. 
\end{proposition}

\begin{proof} The evolution vector field $X_{H,F}$ is characterized by $\contr {X_{H,F}}\omega_{Q}=\dd H+F.$ Observe that 
\begin{align*}
(\mathbb{F} L)^{*}(\contr{X_{H,F}}\omega_{Q})&=(\mathbb{F} L)^{*}(\dd H+F)
&=(\mathbb{F} L)^{*}(\dd H)+(\mathbb{F} L)^{*}(F)
=\dd ((\mathbb{F} L)^{*}H)+(\mathbb{F} L)^{*}F=\dd (E_{L})+F^{L}
\\&
=\contr{X_{L,F^{L}}}\omega_{L}.
\end{align*}
This means that
 \begin{align*}
 \contr {X_{L,F^{L}}}\omega_{L}
 &=(\mathbb{F} L)^{*}(\contr{X_{H,F}}\omega_{Q})
 =\contr{(\mathbb{F} L^{-1})_{*}X_{H,F}}(\mathbb{F} L^{*}\omega_{Q})
 =\contr{(\mathbb{F} L^{-1})_{*}X_{H,F}}\omega_{L}.
 \end{align*} 
 This last implies $X_{L,F^{L}}=(\mathbb{F} L^{-1})_{*}X_{H,F}$, that is, $(\F L)_{*} X_{L,F^{L}}=X_{H,F}.$ \end{proof}

\subsection{Routh reduction for forced mechanical systems}
There exists a large class of systems for which the Lagrangian (or Hamiltonian) does not depend on some of the generalized coordinates. Such coordinates are called \myemph{cyclic} and the corresponding generalized momenta are easily checked to be constants of the motion --see \cite{abraham_foundations_2008, goldstein_classical_1980}. Routh reduction is a classical reduction technique which takes advantage of the conservation laws to define a \myemph{reduced Lagrangian} function, the so-called \myemph{Routhian}, such that, when the conservation of momenta is taken into account, the solutions of the Euler--Lagrange equations for the Routhian are in correspondence with the solutions of the Euler--Lagrange equations for the original Lagrangian. 

Routh reduction can be extended to forced systems as follows \cite{goldstein_classical_1980}.
Suppose that the configuration space is of the form $Q=Q_{1}\times Q_{2}$, where {$Q_1$ is a one-dimensional Lie group. Then, there is a natural Lie group action of $Q_1$ on $Q$, given by left translations on $Q_1$ and the identity on $Q_2$. Note that either $Q_1\simeq\R$ or $Q_1\simeq\mathbb{S}^1$.} We denote an element $q^{i}\in Q$ by $q^{i}=(q^{1},q^j)$, with $q^1\in Q_1$ and $q^j\in Q_{2}$, for $j\in\{2,\ldots,n\}$.

Let $L(q^1,\dot q^1,q^j,\dot q^j)$ be a hyperregular Lagrangian with cyclic coordinate $q^1$, that is, $\displaystyle{\frac{\partial L}{\partial q^1}=0}$ and let $F_{i}$ be a non-conservative force such that $F_{i}$ is independent of $q^1$ for all $i\in \{1, \ldots, n\}$ and $F_{1}(q^2, \ldots ,q^n)=0$. 
{In other words, both $L$ and $F$ are invariant under the action of $Q_1$ on $Q$.}
Fundamental to reduction is the notion of a \myemph{momentum map} $J_L:\T Q\rightarrow \mathfrak{g}^{*}$, which makes explicit the conserved quantities in the system. Here $\mathfrak{g}$ is the Lie algebra associated with the Lie group of symmetries $G$, and $\mathfrak{g}^{*}$ denotes its dual as vector space. In the framework we are considering here, $J_L(q^1,\dot q^1,q^j,\dot q^j)=\displaystyle{\frac{\partial L}{\partial \dot{q}^1}}.$

Fix a value of the momentum $\mu=\displaystyle{\frac{\partial L}{\partial \dot{q}^1}}$. Since $L$ is  hyperregular, the last equation admits an inverse, and allows us to write $\dot{q}^{1}=f(q^2, \ldots ,q^n,\dot{q}^2, \ldots ,\dot{q}^n,\mu)$. Consider the function
	\begin{equation*}\label{eq:Routh}
	R_F^{\mu}(q^j,\dot q^j)=\restr{\left(L-\dot q^1\mu\right)}{\mu}\,,
	\end{equation*} where the notation $\restr{}{\mu}$ means that we have used the relation $\mu=\displaystyle{\frac{\partial L}{\partial \dot{q}^1}}$ to replace all the appearances of $\dot q^1$ in terms of $(q^j,\dot q^j)$ and the parameter $\mu$. The function $R_F^{\mu}$ is called \myemph{Routhian}. Similarly we define the \myemph{reduced force} $F_{\mu}$ as $F_{\mu}\coloneqq \restr{F}{\mu}$.
\begin{remark}	
	The previous definition of Routhian is the classical one considered in the literature as in \cite{goldstein_classical_1980} and \cite{pars_treatise_1965}. A more general notion of Routhian can be given in terms of the so-called \emph{mechanical connection} (see \cite{marsden_lectures_1992}), when $L$ is kinetic, or even more general by using any connection on the principal bundle $Q\to Q/G$. Indeed, let $\mathcal{A}\colon \T Q\to\mathfrak{g}$ be a connection one form, and let us denote by $\mathcal{A}_\mu(\cdot)=\langle \mu,\mathcal{A}(\cdot)\rangle$ the 1-form on $Q$ obtained by contraction with $\mu\in\mathfrak{g}^*$. Then, one can define the Routhian as $R^{\mu}=L-\mathcal{A}_\mu:\T Q\rightarrow\R$. Similarly, one can define the reduced force $F_{\mu}$ by contraction with $\mu\in\mathfrak{g}^{*}$. A review of the unforced Routh reduction in these more general settings can be found in \cite{marsden_lectures_1992} (see also \cite{langerock_routhian_2010}).\end{remark}

\begin{remark}
{A Routhian, as a single function that generates the reduced dynamics, may not always exist, even in the hyperregular case. Instead, one may obtain a family of functions, similar to what happens with the Hamiltonian when applying the Legendre transformation to singular Lagrangians.}
An example of such a case can be found in 
\cite{grabowska_geometry_2019} (see Example $3.4$). Nevertheless in this paper we shall restrict to the case of mechanical Lagrangians (i.e., kinetic minus potential energy), where the last situation cannot happen.  
Indeed, a mechanical Lagrangian with a cyclic coordinate $q^1$ is of the form
$$L=\frac{1}{2}g_{ij}(q^2, \ldots, q^n) \dot{q}^i \dot{q}^j - V(q^2, \ldots, q^n)\, ,$$
where $g_{ij}$ are the components of the Riemannian metric, with $i, j\in \{1, \ldots, n\}$ and $n=\dim Q$. Thus, $\partial L/\partial \dot q^1 = g_{i1} \dot q^i$, and the Routhian is given by
$$R^\mu = \restr{\left(L - \dot q^1 \mu\right)}{\mu} =  \frac{1}{2}\sum_{i, j=2}^n g_{ij} \dot{q}^i \dot{q}^j - V(q^2, \ldots, q^n)\, ,$$
which is regular.

A more geometric and general framework for Routh reduction is presented in \cite{grabowska_geometry_2019}. Following Tulczyjew’s \textit{Weltanschauung} of geometric mechanics, the authors study Routh reduction through symplectic reductions and their generating objects. In this approach, the Routhian, the generator of the reduced dynamics, is not a function but rather a family of sections of an AV-bundle (i.e., an affine bundle modelled on a trivial bundle $M\times \R \to M$).
\end{remark}

If we regard the pair $(R_F^{\mu}, F_{\mu})$ as a new forced Lagrangian system in the variables $(q^j,\dot q^j),\, j\in \{2, \ldots, n\}$, then the solutions of the forced Euler--Lagrange equations for $(R_F^{\mu}, F_{\mu})$ are in correspondence with those for $(L,F)$ when one takes into account the relation $\mu=\displaystyle{\frac{\partial L}{\partial \dot{q}^1}}$. More precisely:
	\begin{itemize}
		\item[(a)] Any solution of the forced Euler--Lagrange equations for $(L, F)$ with momentum $\mu=\displaystyle{\frac{\partial L}{\partial \dot{q}^1}}$ projects onto a solution of the forced Euler--Lagrange equations for $(R_F^{\mu}, F_{\mu})$, given by $\frac{\dd }{dt}\left(\frac{\partial R_F^{\mu}}{\partial\dot{q}^{j}}\right)-\frac{\partial R_F^{\mu}}{\partial q^{j}}=-(F_{\mu})_j$ for  $j\in\{2,\ldots,n\}$. These equations will be referred as \myemph{forced Routh equations} and they induce a vector field $X_{R}^{F}:\T Q_{2}\to \T( \T Q_{2})$ describing the dynamics of the reduced system, called \myemph{Routh vector field}.
		\item[(b)] Conversely, any solution of forced Routh equations for $(R_F^{\mu}, F_{\mu})$ can be lifted to a solution of the forced Euler--Lagrange equations for $(L,F)$ with $\mu=\displaystyle{\frac{\partial L}{\partial \dot{q}^1}}$.
	\end{itemize}

\section{Simple hybrid forced Hamiltonian systems and hybrid constants of motion}\label{section_simple_hybrid}
Roughly speaking, the term \myemph{hybrid system} refers to a dynamical system which exhibits both continuous and discrete time behaviours. In the literature, one finds slightly different definitions of hybrid system depending on the specific class of applications of interest. For simplicity, and following \cite{johnson_simple_1994} and \cite{colombo_time_2018}, we will restrict ourselves to the so-called simple hybrid mechanical systems in Hamiltonian form. 

\myemph{Simple hybrid systems} \cite{johnson_simple_1994} are characterized by the 4-tuple $\mathscr{H}=(D, X, S, \Delta)$, where $D$ is a smooth manifold called the \myemph{domain}, $X$ is a smooth \myemph{vector field} on $D$, $S$ is an embedded submanifold of $D$ with co-dimension $1$ called the \myemph{switching surface}, and $\Delta:S\to D$ is a smooth embedding called the \myemph{impact map}. The submanifold $S$ and the map $\Delta$ are also referred to as the \myemph{guard} and the \myemph{reset map}, respectively. The triple $(D,S,\Delta)$ is called a \myemph{hybrid manifold}. 
{The impact map is assumed to be an embedding to avoid topological pathologies.}

The dynamics associated with a simple hybrid system is described by an autonomous system with impulse effects as in \cite{westervelt_feedback_2018}. We denote by $\Sigma_{\mathscr{H}}$ the \myemph{simple hybrid dynamical system} generated by $\mathscr{H}$, given by 
\begin{equation}
  \label{LHS}\Sigma_{\mathscr{H}}:
  \left\{\begin{array}{ll}
    \dot{\upsilon}(t)=X(\upsilon(t)), & \text{if } \upsilon^{-}(t)\notin{S}, \\ 
    \upsilon^{+}(t)=\Delta(\upsilon^{-}(t)),& \text{if }\upsilon^-(t)\in{S}, 
  \end{array}\right.
\end{equation}
where $\upsilon:I\subset\mathbb{R}\to D$, and $\upsilon^{-}$, $\upsilon^{+}$ denote the states immediately before and after the times when integral curves of $X$ intersect ${S}$ (i.e., pre and post impact of the solution $\upsilon(t)$ with ${S}$), namely $\upsilon^{-}(t)\coloneqq \displaystyle{\lim_{\tau\to t^{-}}}x(\tau)$,\, $\upsilon^{+}(t)\coloneqq \displaystyle{\lim_{\tau\to t^{+}}}x(\tau)$ are the left and right limits of the state trajectory $\upsilon(t)$.


\begin{definition}
	A simple hybrid system $\mathscr{H}=(D, X, {S}, \Delta)$ is said to be a \myemph{simple hybrid forced Hamiltonian system} if it is determined by $\mathscr{H}_{F}\coloneqq (\cT Q, X_{H,F}, {S_H},$ $\Delta_H)$, where $X_{H,F}:\cT Q\to \T( \cT Q)$ is the Hamiltonian forced vector field associated with the forced Hamiltonian system $(H,F)$ (see Subsection \ref{subsec:forced_mechanical_systems}), ${S_H}$ is the switching surface, a submanifold of $\cT Q$ with co-dimension one, and $\Delta_H:{S}_H\to \cT Q$ is the impact map, a smooth embedding.

  The simple hybrid forced dynamical system generated by $\mathscr{H}_{F}$ is given by
  \begin{equation}
    \label{RHDS}\Sigma_{\mathscr{H}_{F}}:
    \left\{\begin{array}{ll}\dot{\upsilon}(t)=X_{H,F}(\upsilon(t)), & \hbox{ if } \upsilon^{-}(t)\notin{S_H},\\ \upsilon^{+}(t)=\Delta_H(\upsilon^{-}(t)),&\hbox{ if } \upsilon^-(t)\in{S_H}, 
    \end{array}\right.
  \end{equation}
  where $\upsilon(t)=(q(t),{p}(t))\in \cT Q $.
\end{definition}

{Mathematically, we will only assume $\Delta_H$ to be a smooth embedding. However, in physical examples one has also to assume that it preserves the base point, i.e., $\Delta_H\circ \pi_Q = \pi_Q$. Otherwise, one could have a mechanical system that teleports in the impacts}

Alternatively, $\Delta_H$ could be described by an impulsive external force appearing only on the instant on the impact (see \cite{ibort_geometric_2001,ibort_mechanical_1997,ibort_geometric_1998,cortes_mechanical_2001} and references therein).




\begin{definition}
A \myemph{simple hybrid forced Lagrangian system} is a simple hybrid system determined by $\mathscr{L}_{F}\coloneqq (\T Q, X_{L,F^L}, {S_L}, \Delta_L)$, where $X_{L,F^L}:\T Q\to \T( \T Q)$ is the forced Lagrangian vector field associated with the forced Lagrangian system $(L,F^L)$, ${S_L}$ the switching surface, a submanifold of $\T Q$ with co-dimension one, and $\Delta_L:{S}_L\to \T Q$ the impact map as defined before.

\end{definition}

\begin{definition}\label{def:flow} A \myemph{hybrid flow} for $\mathscr{H}_{F}$ is a tuple $\chi^{\mathscr{H}_{F}}=(\Lambda,\mathcal{J},\mathscr{C})$, where
	\begin{itemize}
		\item $\Lambda=\{0,1,2, \ldots \}\subseteq \mathbb{N}$ is a countable indexing set,
		\item $\mathcal{J}=\{I_{i}\}_{i\in \Lambda}$ a set of intervals, called \myemph{hybrid intervals}, where $I_{i}=[\tau_{i},\tau_{i+1}]$ if $i, i+1\in \Lambda$; and $I_{N-1}=[\tau_{N-1},\tau_{N}]$ or $[\tau_{N-1},\tau_{N})$ or $[\tau_{N-1},\infty)$ if $|\Lambda|=N$, $N$ finite, with $\tau_{i},\tau_{i+1},\tau_{N}\in \R$ and $\tau_{i}\leq \tau_{i+1}$,
		\item $\mathscr{C}=\{c_{i}\}_{i\in \Lambda}$ is a collection of solutions for the vector field $X_{H,F}$ specifying the continous-time dynamics, i.e., $\dot{c_{i}}=X_{H,F}(c_{i}(t))$ for all $i\in \Lambda$,
		and such that for each $i,i+1\in \Lambda$, (i) $c_{i}(\tau_{i+1})\in S_H$, and (ii) $\Delta_H(c_{i}(\tau_{i+1}))=c_{i+1}(\tau_{i+1})$.
	\end{itemize}
\end{definition}

Similarly, it is possible to define a hybrid flow $\chi^{\mathscr{L}_{F}}$ for a simple hybrid forced Lagrangian system $\mathscr{L}_{F}$. The relation between both hybrid flows is given by the following result

\begin{proposition}\label{Proposition2}
Suppose that $H$ is a hyperregular Hamiltonian.
If $\chi^{\mathscr{H}_{F}}=(\Lambda,\mathcal{J},\mathscr{C})$ is a hybrid flow for $\mathscr{H}_{F}$, $S_L=\F H(S_H)$, and $\Delta_L$ is defined in such a way that $\F H\circ \Delta_H=\Delta_L\circ \restr{\F H}{S}$, then $\chi^{\mathscr{L}_F}=(\Lambda,\mathcal{J},(\F H)(\mathscr{C}))$ with $(\F H)(\mathscr{C})=\{(\F H)(c_{i})\}_{i\in \Lambda}$.
\end{proposition}

\begin{proof}  If $c_i(t)$ is an integral curve of $X_{H,F}$, $\tilde c_i(t)=(\F H\circ c_i)(t)$ is an integral curve for $X_{L,F^L}$. In this way, if we consider a solution $c_0(t)$  with initial value $c_0=(q_0,p_0)$ defined on $[\tau_0,\tau_1]$, then $\tilde c_0(t)$ is a solution with initial value $\tilde c_0=(q_0,\dot{q}_0)$ defined on $[\tau_0,\tau_1]$. Likewise, for a solution $c_1(t)$ defined on $[\tau_1,\tau_2]$, we get a corresponding solution $\tilde c_1(t)$ defined on the same hybrid interval $[\tau_1,\tau_2]$. Proceeding inductively, one finds $\tilde c_i(t)$ defined on  $[\tau_i,\tau_{i+1}]$. 	It only remains to check that $\tilde c_i(t)$ satisfies $\tilde c_i(\tau_{i+1})\in S_{L}$ and $\Delta_{L}(\tilde c_{i}(\tau_{i+1}))=\tilde c_{i+1}(\tau_{i+1})$, but, using the properties of $\F H$,
\begin{itemize}
		\item[(i)] $\tilde c_i(\tau_{i+1})=(\F H\circ c_{i})(\tau_{i+1})=\F H(c_i(\tau_{i+1}))$ and given that $c_i(\tau_{i+1})\in S_H$ then $\tilde c_i(\tau_{i+1})\in S_{L}.$
		\item[(ii)]  $\Delta_L(\tilde c_{i}(\tau_{i+1}))=\Delta_L\circ \F H\circ c_{i}(\tau_{i+1})=\F H\circ \Delta_H\circ c_{i}(\tau_{i+1})=\F H\circ c_{i+1}(\tau_{i+1})=\tilde c_{i+1}(\tau_{i+1})$.
\end{itemize}
\end{proof}



\begin{definition}
Let $\mathscr{H}=(D, X, S, \Delta)$ be a simple hybrid system. A function $f$ on $D$ is called a \myemph{hybrid constant of the motion} if 
\begin{enumerate}[(i)]
\item it is a first integral of $X$, i.e., $X(f)=0$,
\item it is left invariant by the impact map, namely, $f\circ \Delta=f\circ \incl$, where $\incl:S\hookrightarrow D$ denotes the canonical inclusion.
\end{enumerate}
\end{definition}

\section{Generalized hybrid momentum maps and symplectic reduction of simple hybrid forced mechanical systems}\label{section_symp_reduction}

\begin{definition}
Let $G$ be a Lie group and $Q$ a smooth manifold. A \emph{left-action} of $G$ on $Q$ is a smooth map $\psi:G\times Q\to Q$ such that $\psi(e,g)=g$ and $\psi(h,\psi(g,q))=\psi(hg,q)$ for all $g,h\in G$ and $q\in Q$, where $e$ is the identity of the group $G$ and the map $\psi_g:Q\to Q$ given by $\psi_g(q)=\psi(g,q)$ is a diffeomorphism for all $g\in G$.
\end{definition}

\begin{definition}
A Lie group action $\psi\colon G\times Q\to Q$ is said to be a \emph{free} action if it has no fixed points, that is, $\psi_g(q)=q$ implies $g=e$. The Lie group action $\psi$ is said to be a \emph{proper} action if the map
$\tilde{\psi}:G\times Q\to Q\times Q$ given by $\tilde{\psi}(g,q)=(q,\psi(g,q))$, is proper, that is, if $K\subset Q\times Q$ is compact, then $\tilde{\psi}^{-1}(K)$ is compact.
\end{definition}

We recall that for $q\in Q$, the \emph{isotropy} (also known as \emph{stabilizer} or \emph{symmetry}) group of $\psi$ at $q$ is given by $G_q\coloneqq \{g\in G|\psi_g(q)=q\}\subset G.$ Since $\psi_q(g)$ is a continuous map and $G_q=\psi_q^{-1}(q)$ is a closed subgroup, it is a Lie subgroup of $G$. 
{In particular, for each $\mu\in \mathfrak g^*$, we will denote by $G_{\mu}$ be the \myemph{isotropy subgroup} of $G$ in $\mu$ under the co-adjoint action, namely, $G_\mu = \left\{g\in G \mid \hbox{Ad}_g^*\ \mu = \mu  \right\}$.}

Consider a Lie group action $\psi\colon G\times Q\to Q$ of some Lie group $G$ on the manifold $Q$, and assume it is a free and proper action. These conditions ensure that the quotient of a smooth manifold by the action is a smooth manifold \cite{abraham_foundations_2008,lee_quotient_2012}. Let $\mathfrak{g}$ be the Lie algebra of $G$ and $\mathfrak{g}^{*}$ {its dual as a vector space}. There is a natural lift $\psi^{\cT Q}$ of the action $\psi$ to $\cT Q$, the \myemph{cotangent lift}, defined by $(g,(q,p))\mapsto (\cT \psi_{g^{-1}}(q,p))$. In particular  $\psi^{\cT Q}$ enjoys the following properties \cite{abraham_foundations_2008,de_leon_symmetries_2021}:
\begin{itemize} 
	\item[(i)] It preserves the canonical 1-form on $\cT Q$, that is,  $(\psi^{\cT Q}_g)^*\theta_{Q}=\theta_{Q}$ for all $g\in G$.
  Therefore, it is a symplectic action, i.e., $(\psi^{\cT Q}_g)^*\omega_{Q}=\omega_{Q}$ for all $g\in G$.
	\item [(ii)] It admits an $\hbox{Ad}^*$-equivariant\footnote{{Recall that a momentum map $J$ is called $\operatorname{Ad}^\ast$-equivariant if $J(\psi_g^{\cT Q}(x))=\operatorname{Ad}^\ast_{g^{-1}} J(x)$ for each $g\in G$ and each $x\in \cT Q$.}
} momentum map 
	$
	J\colon \cT Q\to \mathfrak{g}^*$ given by $J(\alpha_{q})(\xi)=\left(\contr{\xi_{Q}^{c}}\theta_{Q}\right)(\alpha_{q})=\alpha_q(\xi_Q(q))$ for each $\xi\in \mathfrak{g}$. Here $\xi_{Q}$ is the infinitesimal generator of the action of $\xi\in \mathfrak{g}$ on $Q$ and $\xi_{Q}^{c}$ is the generator of the lifted action on $\cT Q$. {The map $J$ is called the \emph{natural momentum map}.} 
\end{itemize}

\begin{remark} \label{remark_actions_cotangent}
  One could consider a general action $\Phi:G\times \cT Q \to \cT Q$ of $G$ on $\cT Q$, not necessarily lifted from an action of $G$ on $Q$. 
  {
  However, given a $\xi \in \mathfrak{g}$, in order for the natural momentum map $J$ to satisfy $\contr{\xi_{\cT Q}} \omega_Q (\alpha_q) = \dd (J(\alpha_q)(\xi)$ for all $\alpha_q\in \xi$ (i.e., the infinitesimal generator of the $\xi$-action on $\cT Q$ is the Hamiltonian vector field of the function
  $J^\xi\colon \cT Q\ni \alpha_q\mapsto J(\alpha_q)(\xi)\in \R$
  ) it is a necessary and sufficient condition that $\liedv{\xi_{\cT Q}}\theta_Q=0$. For this condition to be verified for every $\xi\in \mathfrak{g}$, we require the action $\Phi$ of $G$ on $\cT Q$ to preserve the $1$-form $\theta_Q$. Then, it is easy to show (e.g., by direct computation in local coordinates) that a necessary and sufficient condition is that $\Phi$ is lifted from an action on $Q$. Moreover, if $\Phi$ is a lifted action, then the natural momentum map $J$ is $\operatorname{Ad}^*$-equivariant momentum map {(see \cite[Theorem 12.4.1]{marsden_introduction_1999})}.} 
  {One may consider more general momentum maps, for which the non-equivariance one-cocycle $\sigma\colon G\to \mathfrak{g}^\ast$, given by $\sigma(g) = J\circ \Phi_g(x) - \operatorname{Ad}^\ast_{g^{-1}}\circ\, J(x)$ for an arbitrary $x\in \cT Q$, is non-trivial. Refer to \cite{ortega_momentum_2004,marsden_hamiltonian_2007} for additional details. This will be extended to the hybrid realm in future works.}
\end{remark}

\begin{definition}
Denote by $\{\phi^X_t\}$ the flow of a vector field $X$ on $Q$. We can define the \myemph{complete lift} $X^c$ of $X$ as the vector field on $\cT Q$ whose flow is the cotangent lift of $\{\phi^X_t\}$ (see \cite{yano_tangent_1973}). In local coordinates, it is given by $X^c=X^i\frac{\partial}{\partial q^i}-p_j\frac{\partial X^j}{\partial q^i}\frac{\partial}{\partial p_i}$.\end{definition}

Let us first introduce the symplectic reduction for the forced Hamiltonian systems $(H,F)$. 
If the Hamiltonian $H$ is $G$-invariant, the subgroup $G_{F}$ of $G$ such that $H$ and $F$ are both $G_{F}$-invariant can be described as follows.
For each $\xi\in\mathfrak{g}$, consider the real-valued function $J^{\xi}:\cT Q\rightarrow\R$ given by $J^{\xi}(\alpha_{q})=\langle J(\alpha_{q}),\xi\rangle,$ that is $J^{\xi}=\contr{\xi_{Q}^{c}}\theta_{Q}$.
 Let $\xi\in\mathfrak{g}$, then $J^{\xi}$ is a conserved quantity for $X_{H,F}$ if and only if $F(\xi_{Q}^{c})=0$ (see \cite{de_leon_symmetries_2021}). {If this holds, then} $\xi$ leaves $F$ invariant if and only if $\contr {\xi_{Q}^{c}}\dd F=0$. In addition, the vector subspace of $\mathfrak{g}$ given by $\mathfrak{g}_{F}=\{\xi\in \mathfrak{g}:F(\xi_{Q}^{c})=0,\,\, \contr {\xi_{Q}^{c}}\dd F=0\}$ is a Lie subalgebra of $\mathfrak{g}$.  Observe that, for each $\xi\in\mathfrak{g}_{F}$, $\xi_{Q}^{c}$ is a symmetry of the forced Hamiltonian system $(H,F)$ (see Appendix \ref{subsection_symmetries}).

Let $G_{F}$ be the Lie group generated by $\mathfrak{g}_{F}$, { and assume that it is a closed Lie subgroup of $G$.} Let $J_{F}:\cT Q\rightarrow \mathfrak{g}_{F}^{*}$ be the reduced momentum map with $\mu\in\mathfrak{g}_{F}^*$ a regular value of $J_{F}$, and let us denote by $(G_{F})_{\mu}$ the isotropy subgroup in $\mu$. Since the $G$-action on $\cT Q$ is free and proper by hypothesis and $\mu$ is a regular value, the $(G_F)_\mu$-action on $J^{-1}(\mu)$ is free. {We will also assume that it is} proper, and thus $J^{-1}(\mu)/(G_F)_\mu$ is a smooth manifold \cite{lee_quotient_2012}.
We have that \cite{de_leon_symmetries_2021}:
\begin{enumerate}[(i)]
\item $J_{F}^{-1}(\mu)$ is a submanifold of $\cT Q$ and $X_{H,F}$ is tangent to it.
\item  The reduced space $M_{\mu}\coloneqq J_{F}^{-1}(\mu)/(G_{F})_{\mu}$ is a symplectic manifold, whose symplectic structure $\omega_{\mu}$ is uniquely determined by  $\pi^{*}_{\mu}\omega_{\mu}=\incl^{*}_{\mu}\omega_{Q},$ where $\pi_{\mu}:J_{F}^{-1}(\mu)\rightarrow M_{\mu}$ and $\incl_{\mu}:J_{F}^{-1}(\mu)\hookrightarrow \cT Q$ denote the canonical projection and the canonical inclusion, respectively. 
 \item $H$ induces a reduced function $H_\mu:M_{\mu}\rightarrow\R$ defined by $H_\mu\circ\pi_{\mu}=H\circ \contr {\mu}.$
 \item $F$ induces a reduced 1-form $F_{\mu}$ on $M_{\mu},$ uniquely determined by  $\pi^{*}_{\mu}F_{\mu}=\incl^{*}_{\mu}F$. 
 \item The forced Hamiltonian vector field $X_{H,F}$ projects onto $X_{H_\mu, F_\mu}$.
\end{enumerate}

\begin{remark} \label{remark_symmetries_reduction}
In order to obtain a reduced Hamiltonian function $H_\mu$ and a reduced external force $F_\mu$, both $H$ and $F$ need to be, independently, $G_F$-invariant. The conditions that each $\xi_Q^c$ has to satisfy for this to occur are stronger than the ones required for being a symmetry of the forced Hamiltonian (see Subsection \ref{subsection_symmetries}). As a matter of fact, we can weaken this requirements and reduce $\alpha_{H, F}\coloneqq \dd H+ F$ instead of $H$ and $F$ separately. Suppose that $\alpha_{H,F}$ is $\psi^{\cT Q}$-invariant, i.e., $\alpha_{H,F}(\xi_Q^c)=0$ and $\liedv{\xi_Q^c}\alpha_{H,F} = 0$ for every $\xi \in \mathfrak g$. In particular, $\xi_Q$ is a symmetry of the forced Hamiltonian for every $\xi\in \mathfrak g$ (since the action $\psi^{\cT Q}$ leaves $\theta_Q$ invariant, $\liedv{\xi_Q^c} \theta_Q= 0$). Let $G_\mu$ be the isotropy group of $G$ in $\mu$, where $\mu\in\mathfrak{g}^*$ is a regular value of $J$. {Then, $\alpha_{H,F}$ induces a reduced 1-form $\alpha_{H,F}^{\mu}$ on $M_{\mu},$ uniquely determined by  $\pi^{*}_{\mu}\alpha_{H,F}^{\mu}=\incl^{*}_{\mu}\alpha_{H,F}$; and the forced Hamiltonian vector field $X_{H,F}=X_{\alpha_{H,F}}$ projects onto $X_{\alpha_{H,F}^\mu}$, where $\contr{X_{\alpha_{H,F}^\mu}}\omega_\mu=\alpha_{H,F}^\mu$.}
\end{remark}

Next, we extend the symplectic reduction for forced Hamiltonian systems to simple hybrid forced Hamiltonian systems with symmetries. Consider a simple hybrid forced Hamiltonian system $\mathscr{H}_{F}=(\cT Q,X_{H,F},S_H,\Delta_H)$. To perform a hybrid reduction one needs to impose some compatibility conditions between the action and the hybrid system (see \cite{ames_hybrid_2006-1,ames_hybrid_2006}). By a \myemph{hybrid action} on the simple hybrid forced Hamiltonian system $\mathscr{H}_F$  we mean a Lie group  action $\psi\colon G\times Q\to Q$ such that
\begin{itemize}
	\item $H$ is invariant under $\psi^{\cT Q}$, i.e. $H\circ \psi^{\cT Q}=H$,
	\item $\psi^{\cT Q}$ restricts to an action of $G$ on $S_H$,
	\item $\Delta_H$ is equivariant with respect to the previous action, namely $$\Delta_H\circ \restr{\psi^{\cT Q}_g}{S_H}=\psi^{\cT Q}_g\circ \Delta_H.$$
\end{itemize}
{The second requirement implies that $\psi$ restricts to an action of $G$ on $\pi_Q(S_H)$. The converse is false, i.e., $\psi$ may restrict to an action of $G$ on $\pi_Q(S_H)$ without $\psi^{\cT Q}$ restricting to an action of $G$ on $S_H$. A counterexample is $G=Q$ acting on itself by left multiplication, so that $\cT Q = G \times \mathfrak{g}^\ast$ and $S_H = G \times V$, with $V$ a $1$-codimensional vector subspace of $\mathfrak{g}^\ast$.
}

\begin{definition}
A momentum map $J$ will be called a \myemph{generalized hybrid momentum map} for $\mathscr{H}_F$ if, for each connected component $C\subseteq S$ and for each regular value $\mu_-$ of $J$,
\begin{equation}
  \Delta_H \left(\restr{J}{C}^{-1}(\mu_-)  \right) \subset J^{-1}(\mu_+),
  \label{generalized_hybrid_momentum}
\end{equation}
for some regular value $\mu_+$. In other words, for every point in the connected component $C$ of the switching surface $S$ such that the momentum before the impact takes a value of $\mu_-$, the momentum will take a value $\mu_+$ after the impact. That is, the switching map  translates the dynamics from one level set of the momentum map into another.

A generalized hybrid momentum map is called a \myemph{hybrid momentum map} if $\Delta_H$ preserves the momentum map.
 In other words, $J$ is a {hybrid momentum map} if the diagram
\begin{equation}\label{diag1}
\begin{tikzcd}[column sep=1.5cm, row
    sep=1.2cm]
& \mathfrak{g}^* &\\
\cT Q \arrow[ur,"J"] & S_H \arrow[u,"\restr{J}{S_H}"] \arrow[l,swap,hook'] \arrow[r,"\Delta_H"] & \cT Q  \arrow[ul, swap,"J"]
\end{tikzcd} 
\end{equation}
commutes (see \cite{ames_hybrid_2006,ames_hybrid_2006-1}). Note that  If $J$ is a hybrid momentum map, then $J^\xi$ is a hybrid constant of the motion for each $\xi \in \mathfrak g$. Note that this is not the case for a generalized hybrid momentum map.
\end{definition}

Consider $\mathscr{H}_{F}=(\cT Q,X_{H,F}, S_H,\Delta_H)$ equipped with a hybrid action $\psi$ such that $H$ and $F$ are $G_F$-invariant with $G_{F}\subset G$ being the Lie subgroup generated by $\mathfrak{g}_{F}$ and $J_{F}:\cT Q\rightarrow \mathfrak{g}_{F}^{*}$ the reduced momentum map, which is also assumed to be a generalized hybrid momentum map. 
Then, for each $\xi\in \mathfrak g$, $J_F^\xi=\left\langle J_F, \xi  \right\rangle$ is a hybrid constant of the motion. 

Let $\mu_-, \mu_+\in\mathfrak{g}_{F}^*$ be two \myemph{regular hybrid values} of $J_{F},$ which means that they are regular values of both $J_{F}$ and $\restr{J_{F}}{S_H}$. When we combine this definition with the condition \eqref{generalized_hybrid_momentum}, we obtain that the following diagram commutes
\[
\begin{tikzcd}[column sep=1.3cm, row
sep=1.2cm]
J_F^{-1}(\mu_-)\arrow[d,hook'] &  \restr{J_F}{S_H}^{-1}(\mu_-)\arrow[r,"\restr{\Delta_H}{J(\mu_-)^{-1}}"] \arrow[l,swap,hook']\arrow[d,hook'] & J_F^{-1}(\mu_+)\arrow[d,hook']\\
\cT Q  & S_H \arrow[l,swap,hook'] \arrow[r,"\Delta_H"] &  \cT Q  
\end{tikzcd} 
\] where $J_F^{-1}(\mu)$ and $\restr{J_F}{S_H}^{-1}(\mu)$ are embedded submanifolds of $\cT Q$ and $S_H$, respectively. The hook arrows $\hookrightarrow$ in the diagram denote the corresponding canonical inclusions.

\begin{proposition}\label{Proposition:isotropy_subgroups}
Let $\mathscr{H}_{F}=(\cT Q,X_{H,F}, S_H,\Delta_H)$ be a hybrid forced Hamiltonian system, and let $\psi:G\times Q \to Q$ be a Lie group action of a connected Lie group $G$ on $Q$. 
If $\Delta_H$ is equivariant with respect to $\psi^{\cT Q}$, and
 $\mu_-,\ \mu_+$ are regular values of $J$ such that $\Delta_H \left(\restr{J}{S_H}^{-1}(\mu_-)  \right)\subset J^{-1}(\mu_+)$, then $G_{\mu_-}=G_{\mu_+}$.
\end{proposition}

\begin{proof}  Let $g\in G_{\mu_-}$. Then,
\begin{equation}
\begin{aligned}
  J \circ \Delta_H \left(\restr{J}{S_H}^{-1}(\mu_-)  \right)
  &= J \circ \Delta_H\circ \psi_g^{\cT Q} \left(\restr{J}{S_H}^{-1}(\mu_-)  \right)
  \\&
  = J \circ \psi_g^{\cT Q} \circ \Delta_H \left(\restr{J}{S_H}^{-1}(\mu_-)  \right)
  \\&
  = \hbox{Ad}_{g^{-1}}^* \circ J \circ \Delta_H \left(\restr{J}{S_H}^{-1}(\mu_-)  \right),
\end{aligned}
\end{equation}
where we have used the equivariance of $J$ and $\Delta_H$, so $g\in G_{\mu_+}$, and hence $G_{\mu_-}$ is a Lie subgroup of $G_{\mu_+}$. 

Now, observe that $G_\mu$ has the same dimension, for each $\mu\in \mathfrak g^*$. Therefore, the identity components of $G_{\mu_-}$ and $G_{\mu_+}$ coincide. If we assume that $G$ is connected, $G_{\mu_-}$ and $G_{\mu_+}$ are equal to their identity components, so $G_{\mu_-}=G_{\mu_+}$. \end{proof}



\begin{theorem}\label{theorem_reduction_autonomous}
Let $\mathscr{H}_{F}=(\cT Q,X_{H,F}, S_H,\Delta_H)$ be a hybrid forced Hamiltonian system. Let $\psi:G\times Q \to Q$ be a hybrid action of a connected Lie group $G$ on $Q$.
Suppose that $H$ and $F$ are $G_F$-invariant and assume that $J_F$ is a generalized hybrid momentum map.
Consider a sequence $\left\{\mu_i  \right\}$ of regular values of $J_F$, such that $\Delta_H \left(\restr{J}{S_H}^{-1}(\mu_i)  \right)\subset J^{-1}(\mu_{i+1})$.
Let $(G_{F})_{\mu_i}=(G_{F})_{\mu_0}$ be the isotropy subgroup in $\mu_i$
 under the co-adjoint action. Then,
\begin{enumerate}[(i)]
\item $J_{F}^{-1}(\mu_i)$ is a submanifold of $\cT Q$ and $X_{H,F}$ is tangent to it.
\item  The reduced space $M_{\mu_i}\coloneqq J_{F}^{-1}(\mu_i)/(G_{F})_{\mu_0}$ is a symplectic manifold, whose symplectic structure $\omega_{\mu_i}$ is uniquely determined by  $\pi^{*}_{\mu_i}\omega_{\mu_i}=\incl^{*}_{\mu_i}\omega_{Q},$ where $\pi_{\mu_i}:J_{F}^{-1}(\mu_i)\rightarrow M_{\mu_i}$ and $\contr {\mu_i}:J_{F}^{-1}(\mu_i)\hookrightarrow \cT Q$ denote the canonical projection and the  canonical inclusion, respectively. 
 \item 
 $(H,F)$ induces a reduced forced Hamiltonian system $(H_{\mu_i}, F_{\mu_i})$ on $M_{\mu_i}$, given by $H_{\mu_i}\circ\pi_{\mu_i}=H\circ \contr {\mu_i}$ and $\pi^{*}_{\mu_i}F_{\mu_i}=\incl^{*}_{\mu_i}F$. Moreover, the forced Hamiltonian vector field $X_{H,F}$ projects onto $X_{H_{\mu_i},F_{\mu_i}}$.
 \item $\restr{J_{F}}{S_H}^{-1}(\mu_i)\subset S_H$ reduces to a submanifold of the reduced space
   $(S_H)_{\mu_i}\subset J_{F}^{-1}(\mu_i)/(G_{F})_{\mu_0}.$ 
 \item 
 $\Delta_{H\mid J^{-1}(\mu_i)}$ reduces to a map $(\Delta_H)_{\mu_i}:(S_H)_{\mu_i} \rightarrow J_{F}^{-1}(\mu_{i+1})/(G_{F})_{\mu_0}.$
\end{enumerate}
{Therefore, after the reduction procedure, we get a sequence of reduced simple hybrid forced Hamiltonian systems $\left\{\mathscr{H}_ {F}^{\mu_i}  \right\}$, where 
$\mathscr{H}_ {F}^{\mu_i}=(J_{F}^{-1}(\mu_i)/(G_{F})_{\mu_0},$ $X_{H_{\mu_i},F_{\mu_i}},(S_H)_{\mu_i},(\Delta_H)_{\mu_i})$.}

The reduction scheme is summarized in the following commutative diagram:
\begin{center}
\begin{tikzcd}
\cdots \arrow[r] & J_F^{-1}(\mu_i) \arrow[dd]        &  & \restr{J_F}{S_H}^{-1}(\mu_i) \arrow[dd] \arrow[rr, "\Delta_{H\mid J^{-1}(\mu_i)}"] \arrow[ll, hook] &  & J_F^{-1}(\mu_{i+1}) \arrow[dd]        & \cdots \arrow[l, hook] \\
                 & {} \arrow[d]                      &  &                                                                                               &  &                                       &                        \\
\cdots \arrow[r] & \frac{J_F^{-1}(\mu_i)}{G_{\mu_0}} &  & \left(S_H\right)_{\mu_i} \arrow[rr, "\left(\Delta_H\right)_{\mu_i}"] \arrow[ll, hook]         &  & \frac{J_F^{-1}(\mu_{i+1})}{G_{\mu_0}} & \cdots \arrow[l, hook]
\end{tikzcd}

\end{center}

\end{theorem}

\begin{proof} 
{The fundamental idea is that, by Proposition~\ref{Proposition:isotropy_subgroups}, we can quotient all the regular level sets of $J_F$ and $\restr{J_{F}}{S_H}$ by the same isotropy subgroup $(G_{F})_{\mu_0}$.}

See \cite{abraham_foundations_2008,ortega_momentum_2004} for a proof of the first two assertions. The third statement was proven in \cite{de_leon_symmetries_2021}. 

Since the $(G_{F})_{\mu_0}$-action restricts to a free and proper action on $S_H$, $(S_H)_{\mu_i}=\restr{J_{F}}{S_H}^{-1}(\mu_i)/(G_{F})_{\mu_0}$ is 
a smooth manifold. Clearly, it is
 a submanifold of\\ $J_{F}^{-1}(\mu_i)/(G_{F})_{\mu_0}$. Since $\Delta_H$ is equivariant, it induces an embedding $(\Delta_H)_{\mu_i}:(S_H)_{\mu_i} \rightarrow J_{F}^{-1}(\mu_{i+1})/(G_{F})_{\mu_0}$.\end{proof}

The reduction picture in the Lagrangian side can now be obtained from the Hamiltonian one by adapting the scheme developed in~\cite{langerock_routhian_2010}. In the same fashion as in the Hamiltonian side, by a \myemph{hybrid action} on the simple hybrid Lagrangian system $\mathscr{L}_F = (\T Q, X_{L,F^L}, {S_L}, \Delta_L)$ we mean a Lie group action $\psi\colon G\times Q\to Q$ such that
\begin{itemize}
	\item $L$ is invariant under $\psi^{\T Q}$, i.e. $L\circ \psi^{\T Q}=L$,
	\item $\psi^{\T Q}$ restricts to an action of $G$ on $S_L$,
	\item $\Delta_L$ is equivariant with respect to the previous action, namely $\Delta_L\circ \restr{\psi^{ \T Q}_g}{S_L}=\psi^{\T Q}_g\circ \Delta_L$,
\end{itemize} where $\psi^{\T Q}$ is the tangent lift of the action $\psi$ to $\T Q$, defined by $(g,(q,\dot q))\mapsto (\T\psi_{g}(q,\dot q))$.

The key idea is that, since $\psi^{\T Q}$ is a hybrid action under which $L$ is invariant, the Legendre transformation $\mathbb{F} L$ is a diffeomorphism such that:
\begin{itemize}
	\item it is equivariant with respect to $\psi^{\T Q}$ and $\psi^{\cT Q}$,
	\item it preserves the level sets of the momentum map, that is, $\mathbb{F} L ((J_L)_{F}^{-1}({\mu_i}))=J_{F}^{-1}({\mu_i}),$
	\item it relates the symplectic structures, that is, $(\mathbb{F} L)^*\omega_{Q}=\omega_L$, meaning that, $\mathbb{F} L$ is a \myemph{symplectomorphism}.
\end{itemize}
It follows that the map $\mathbb{F} L$ reduces to a symplectomorphism $(\mathbb{F} L)_{\text{red}}$ between the reduced spaces. Consider the Lie subalgebra $\mathfrak{g}_{F^L}=\{\xi\in \mathfrak{g}:F^L(\xi_{Q}^{c})=0,\,\, \contr {\xi_{Q}^{c}}\dd F^L=0\}$ of $\mathfrak g$, and let $G_{F^L}$ be the Lie subgroup it generates. 
 Then, the following diagram commutes:
\begin{center}
\begin{tikzcd}[column sep=1.4cm, row sep=1.2cm]
(\T Q,S_{L},\Delta_{L})  \arrow[d,swap,"\text{Red.}"] \arrow[r,"\mathbb{F} L"] & (\cT Q,S_H,\Delta_H) \arrow[d,"\text{Red.}"]\\
(\mathcal{M}^{L}_{\mu_0},(S_{L})_{\mu_i},(\Delta_{L})_{\mu_i}) \arrow[r,"(\mathbb{F} L)_{\text{red}}"] & (\mathcal{M}^{H}_{\mu_0},(S_H)_{\mu_i},{(\Delta_H)_{\mu_i}})
\end{tikzcd}\, ,
\end{center}
where we have used the notation $\mathcal{M}^{L}_{\mu_0}\coloneqq (J_L)_{F}^{-1}(\mu_i)/(G_{F^{L}})_{\mu_0}$ and $\mathcal{M}^{H}_{\mu_0}\coloneqq J_{F}^{-1}(\mu_i)/(G_{F})_{\mu_0}$.

\begin{lemma}\label{lemma1}
If $(L, F^L)$ is the Lagrangian counterpart of $(H, F)$ (i.e., {the Lagrangian energy is $E_L = H \circ \F L$ and the Lagrangian external force is $F^L = \F L^* F$}), then $G_{F^L}=G_F$.\end{lemma}
\begin{proof} For any $g\in G_{F}$, 
\begin{equation}
\begin{aligned}
  \left( \psi_g^{\T Q}  \right)^* F^L
  &=  \left( \psi_g^{\T Q}  \right)^* \circ \F L^* F= \left( \F L \circ \psi_g^{\T Q}  \right)^* F
  = \left( \psi_g^{\cT Q} \circ \F L  \right)^* F= \F L^* \circ \left( \psi_g^{\cT Q}  \right)^* F
  = \F L^* F 
  = F^L,
\end{aligned}
\end{equation}
where we have used the equivariance of $\F L$, so $g \in G_{F^L}$. Similarly one can show that, for any $g \in G_{F^L}$, $g\in G_F$.\end{proof}


In the following we make use of a principal connection on the bundle $Q\to Q/G$ to make some further identifications. In the case of a mechanical Lagrangian, a natural choice is the so-called mechanical connection \cite{marsden_lectures_1992}.
 Let $\mathcal{A}\colon \T Q\to\mathfrak{g}$ be a connection one form, and let us denote by $\mathcal{A}_{\mu_i}(\cdot)=\langle {\mu_i},\mathcal{A}(\cdot)\rangle$ the 1-form on $Q$ obtained by contraction with ${\mu_i}\in\mathfrak{g}^*$. Building on the well-known results on cotangent bundle reduction \cite{marsden_hamiltonian_2007} it is possible to show that there is an identification
\begin{equation}\label{eq:ide}
\begin{aligned}
J_{F}^{-1}({\mu_i})/(G_{F})_{\mu_0} &\simeq  \cT (Q/G_F)\times_{Q/G_F}Q/(G_{F})_{\mu_0} \\
&{
=\left\{(x,y)\in \cT (Q/G_F)\times Q/(G_{F})_{\mu_0}\mid \pi_{Q/G_F}(x) = \pi(y) \right\}\, ,
}
\end{aligned}
\end{equation}
{where $\pi_{Q/G_F} \colon \cT (Q/G_F)\to Q/G_F$ and $\pi \colon Q/(G_{F})_{\mu_0}\to Q/G_F$ are the natural projections.
This identification is a symplectomorphism when the space on the right-hand side of~\eqref{eq:ide} is endowed with the symplectic structure ${\rm pr}_1^*\omega_{Q/G_F}+{\rm pr}_2^*\mathcal{B}_{\mu_i}$, where $\pr_1\colon \cT (Q/G_F)\times_{Q/G_F}Q/(G_{F})_{\mu_0} \to \cT (Q/G_F)$ and $\pr_2\colon \cT (Q/G_F)\times_{Q/G_F}Q/(G_{F})_{\mu_0} \to \cT (Q/G)$ are the natural projections, $\omega_{Q/G_F}$ is the canonical symplectic form of the cotangent bundle $\cT (Q/G_F)$,
and $\mathcal{B}_{\mu_i}$ is the so-called magnetic term, obtained from the reduction of $d\mathcal{A}_{\mu_i}$ to $Q/(G_F)_{\mu_0}$.} For details, see~\cite{langerock_routhian_2010,marsden_hamiltonian_2007}.

For the reduction in the Lagrangian side, one needs an additional regularity condition, sometimes referred to as \myemph{$G$-regularity}. Precisely, one has the following definition~\cite{langerock_routh_2011}.


\begin{definition} Consider a $G_{F^L}$-invariant forced Lagrangian system $(L,F^{L})$ on $\T Q$ (i.e., $L\circ\psi_g^{\T Q}=L$ and $(\psi_g^{\T Q})^* F^L=F^L$ for every $g\in G$) and let $\xi^{c}_Q$ be the infinitesimal generator for the associated lifted action. Then, $(L,F^{L})$ is said to be \myemph{$G_{F^{L}}$-regular} if, for each $v_q\in \T Q$, the map $(\mathcal{J}_L)_{F^L}^{v_q}:\mathfrak{g}_F\to \mathfrak{g}_F^*$,\,
$\xi\mapsto (J_L)_{F^L}\left(v_q + \xi^{c}_Q(q)\right)$ is a diffeomorphism.
\end{definition}


Essentially, the $G_{F^L}$-regularity demands regularity ``with respect to the subgroup variables''. Hereinafter, the pair $(L,F^{L})$ will be assumed to be $G_{F^L}$-regular, so that there is an identification $$(J_L)_{F}^{-1}({\mu_i})/(G_{F^L})_{\mu_0}\simeq  \left(\T( Q/G)\times_{Q/G}Q/(G_{F^L})_{\mu_0}\right)$$ 
(see \cite{garcia-torano_andres_aspects_2014} for instance).

The reduced dynamics on this space can be interpreted as the Lagrangian dynamics of some regular Lagrangian subjected to a gyroscopic force (arising from the magnetic term) by working in the more general class of \myemph{magnetic Lagrangians}~\cite{langerock_routh_2012}, which in the present situation need to be extended in order to include external forces. Magnetic Lagrangian systems are a broad family of Lagrangian systems on which the Lagrangian function might be independent of some of the velocities. A force term given by a 2-form can also appear in these systems. Since the Routh reduction yields, in general, a reduced system which is not a standard Lagrangian system, magnetic Lagrangian systems provide a quite convenient framework for carrying out Routh reduction. The extension to magnetic Lagrangian systems allows to carry out Routh reduction by stages~\cite{langerock_routh_2011}. The role of the reduced Lagrangian function is played by the \myemph{Routhian}\footnote{Note the difference with the Hamiltonian reduction.}, which is defined as (the reduction of) the $(G_{F})_{\mu_0}$-invariant function $R^{\mu_i}_F=L-\mathcal{A}_{\mu_i}$ restricted to $(J_L)_F^{-1}({\mu_i})$. 
The next diagram summarizes the situation:

\begin{center}
\begin{tikzcd}[column sep=1.2cm, row
sep=1.2cm]
\T Q  \arrow[d,swap,"\text{Red.}"] \arrow[r,"\mathbb{F} L"] &  \cT Q \arrow[d,"\text{Red.}"]\\
  \left(\T( Q/G)\times_{Q/G}Q/(G_{F^L})_{\mu_0}\right) \arrow[r,"\mathbb{F} R^{\mu_i}_F"] &   \left(\cT (Q/G)\times_{Q/G}Q/(G_{F})_{\mu_0}\right)
\end{tikzcd}
\end{center}




\subsection{A particular case: cyclic coordinates}\label{cyclic}

When $(L, F^L)$ is $\mathbb{S}^1$-invariant, one recovers the classical notion of a cyclic coordinate (the case $G=\mathbb{R}$ is analogous; if $G$ is a product of $\mathbb{S}^1$ or $\R$ one can iterate the procedure). Since $G=\mathbb{S}^1$ is abelian, $G_{\mu_i}=G$ for every $\mu_i \in \mathfrak g^*$. The reduced space  $(J_L)_{F^L}^{-1}({\mu_i})/(G_{F})_{\mu_0}$ can be identified with $ \T( Q/\mathbb{S}^1)$. Similarly, the reduced switching surface $(S_L)_{\mu_i}$ can be identified with a submanifold of $\T( Q/\mathbb{S}^1)$, and the impact map can be identified with a map $(\Delta_L)_{\mu_i}\colon (S_L)_{\mu_i}\to \T( Q/\mathbb{S}^1)$.


If the forced Lagrangian system $(L,F^L)$ has a cyclic coordinate $\theta$, i.e., $L$ is a function of the form $L(\dot{\theta},x,\dot{x})$, and $F$ is of the form $F(\dot{\theta},x,\dot{x}) = F_{x}(\dot{\theta},x,\dot{x}) \dd x$, the conservation of the momentum map $(J_L)_{F}={\mu_i}$ reads $\frac{\partial L}{\partial \dot{\theta}}={\mu_i}$. This relation can be used to write $\dot{\theta}$ as a function of the remaining --non cyclic-- coordinates and their velocities, and the fixed regular value of the momentum map ${\mu_i}$, namely, $\dot \theta =\dot \theta(x, \dot x, \mu_i)$.
It is worth noting that this is the stage where the
$G_F$-regularity of $(L, F^L)$ is used, in order to guarantee that $\dot{\theta}$ can be expressed in terms of $x$, $\dot{x}$ and ${\mu_i}$. If the connection on the bundle $Q\rightarrow Q/\mathbb{S}^1=M$ is chosen to be the canonical flat connection, then the Routhian and the reduced external force can be written as
\begin{align} \label{eqq5}
  R_{F}^{\mu_i}(x,\dot x)&=\restr{\left[L(\dot{\theta},x,\dot x)-{\mu_i}\dot{\theta} \right]}{\dot{\theta}=\dot{\theta}(x,\dot x,{\mu_i})},\\  F^L_{\mu_i}(x,\dot x)&=\restr{F^L(\dot{\theta},x,\dot{x})}{\dot{\theta}=\dot{\theta}(x,\dot x,{\mu_i})}.   
\end{align}
 Here the notation means that $\dot{\theta}$ is expressed as a function of $(x,\dot x,{\mu_i})$. Observe that~\eqref{eqq5} coincides with the classical definition of the Routhian~\cite{pars_treatise_1965}. Moreover, since the connection is flat, no magnetic terms appear in the reduced dynamics.

The value of the momentum map will, in general, be modified in the (non-elastic) collisions with the switching surface. Therefore, the reduced Hamiltonian $H_{\mu_i}$ and the reduced external force $F_{\mu_i}$ will have to be defined in each $I_i$, where $\mathcal{J}=\{I_{i}\}_{i\in \Lambda}$ is the hybrid interval, and they will depend on  the value of the momentum $\mu_i$ after the collision at time $\tau_i$. It is worth noting that this also affects the way the impact map $\Delta_H$ is reduced. 

Let us denote: (1) $\mu_i$ the momentum of the system in $I_i=[\tau_i,\tau_{i+1}]$, (2) $(\Delta_H)_{\mu_i}$ the reduction of $\Delta_{H\mid J^{-1}(\mu_i)}$, and (3) ${(S_H)}_{\mu_i}$ the reduction of ${S_H}$, so there is a sequence of reduced simple hybrid forced Hamiltonian systems:
\begin{center}
\begin{tikzcd}[column sep=.5cm, row
sep=.7cm]
{[\tau_0,\tau_1]} \arrow[d,swap,"\text{Coll.}"]\arrow[r,"\text{Red.}"] & (\cT (Q/\mathbb{S}^1)\times_{Q/\mathbb{S}^1}Q/(G_{F})_{\mu_0},X_{H_{\mu_{0}},F_{\mu_{0}}},(S_H)_{\mu_0},(\Delta_H)_{\mu_0}) \arrow[d,"\text{Coll.}"]\\
{[\tau_1,\tau_2]} \arrow[d,swap,"\text{Coll.}"]\arrow[r,"\text{Red.}"] & (\cT (Q/\mathbb{S}^1)\times_{Q/\mathbb{S}^1}Q/(G_{F})_{\mu_0},X_{H_{\mu_{1}},F_{\mu_{1}}},(S_H)_{\mu_1},(\Delta_H)_{\mu_1}) \arrow[d,"\text{Coll.}"]\\
(\dots) \arrow[r,"\text{Red.}"] & (\dots)
\end{tikzcd} 
\end{center}
Here ``Coll.'' and ``Red.'' stand for collision and reduction, respectively.

Since the momentum, generally, changes with the collisions the reconstruction procedure will be more challenging. In order to make use a reduced solution to reconstruct the original dynamics, the reduced hybrid data have to be computed after each collision. That is, once the reduced solution for the time interval between two collision events, say between $t=\tau_{i} $ and $t=\tau_{i+1},$ has been obtained, this solution has to be reconstructed to obtain the new momentum after the collision at $\tau_{i+1}$. After that, this new momentum has to be used in order to build a new reduced hybrid system {$\mathscr{H}_ {F}^{\mu_{i+1}}=(J_{F}^{-1}(\mu_{i+1})/(G_{F})_{\mu_0},$ $X_{H_{\mu_{i+1}},F_{\mu_i}},(S_H)_{\mu_{i+1}},(\Delta_H)_{\mu_{i+1}})$} whose solution should be obtained until the next collision event at $\tau_{i+2}$, and so forth. As usual,
in order to reconstruct the hybrid flow from the reduced hybrid flow, one has to integrate the regular value at each stage in the previous diagram, using the solution of the reduced simple hybrid forced Hamiltonian system. Essentially, this is tantamount to imposing the momentum constraint on the reconstructed solution.

{What can make the reconstruction of hybrid flow significantly more involved than the usual reconstruction of a (forced) Hamiltonian flow is the dependence of the (forced) Hamiltonian vector field $X_{H_{\mu_{i+1}},F_{\mu_i}}$, the switching surface $(S_H)_{\mu_{i+1}}$, and the impact map $(\Delta_H)_{\mu_{i+1}}$ on the value $\mu_i$ of the momentum map, which changes on each impact. However, there are some cases in which the changes of these objects with the impacts is not complicated (see Example~\ref{example_disk}).}


More precisely, suppose that $\chi^{\mathscr{L}_{F}^{\mu_i}}(c_0)=(\Lambda,\mathcal{J},\mathcal{C}^{\mu_i})$ is a hybrid flow of $\mathscr{L}_{F}^{\mu_i}$. Then we can construct a hybrid flow $\mathscr{L}_{F}(c_0(\tau_0))=(\Lambda,\mathcal{J},\mathcal{C})$ of $\mathscr{L}_{F}$ by constructing the flow recursively between two collisions. Writing $c_i^{\mu_i}(t)=(x_{\mu_i},\dot{x}_{\mu_i})$, we define $c_{i}(t)=(x_{\mu_i},\dot{x}_{\mu_i},\theta_{\mu_i},\dot{\theta}_{\mu_i}))$ recursively as follows. Assume that we have a mechanical Lagrangian of the form $L(x, \dot x, \dot \theta) = \dot q^T M \dot q - V(q)$, where $\displaystyle{\dot q = \begin{pmatrix}
\dot x\\ \dot \theta
\end{pmatrix}}$ and the mass matrix is $\displaystyle{
W=\begin{pmatrix}
  W_x(x) & W^{T}_{\theta,x}(x)\\ 
   W_{\theta,x}(x) & W_{\theta}(x)
\end{pmatrix}}$.
First note that $(J_L)_F(x,\dot{x},\theta,\dot{\theta})=\frac{\partial L}{\partial\dot{\theta}}(x,\dot{x},\theta,\dot{\theta})=W_{\theta,x}(x)\dot{x}+W_{\theta}(x)\dot{\theta}$. Then, it is easy to see that
\begin{align}
    \dot{\theta}_{\mu_i}(t)=&W_{\theta}^{-1}(\theta_{\mu_i}(t))(\mu_i-W_{\theta,x}(x_{\mu_i})\dot{x}_{\mu_i}(t)),\label{eqqq1}\\
    \theta_{\mu_i}(t)=&(\Delta_H^{\theta})_{\mu_i}(c_{i-1}^{\mu_{i-1}}(\tau_i))+\int_{\tau_i}^{t-\tau_i}\dot{\theta}_{\mu_i}(s)\, \dd s,\label{eqqq2}
\end{align}
where $t\in[\tau_i,\tau_{i+1}]$ and $(\Delta_H^{\theta})_{\mu_i}(c_{i-1}^{\mu_{i-1}}(\tau_i))$ is the $\theta$-component of $(\Delta_H)_{\mu_i}(c_{i-1}^{\mu_{i-1}}(\tau_i))$. Note that, at each step, one has to reconstruct with the corresponding momenta $\mu_i$ in equation \eqref{eqqq1} and reduce again the dynamics after the collision with a new momenta 
$\mu_{i+1}$ as conserved quantity.

{An interesting question that we defer to future work is how the long-time dynamics look after an arbitrarily large number of impacts have occurred. In particular, it is natural to consider long-time stability of hybrid systems. It can be studied via the Poincar\'e map \cite{goodman_existence_2020}, some of them - depending the dimensions of the systems - with Poincar\'e-Bendixon Theorem \cite{william2020poincare} and many other techniques (see \cite{haddad2006impulsive} for instance).}

\begin{remark}
     If the momentum map is a hybrid momentum map, the reduction of a simple hybrid forced Hamiltonian system $\mathscr{H}_{F}=(\cT Q,X_{H,F}, S_H,\Delta_H)$, with initial value of the momentum map $\mu_0$, yields a single reduced simple hybrid forced Hamiltonian system $$\mathscr{H}_ {F}^{\mu_0}=(J_{F}^{-1}(\mu_0)/(G_{F})_{\mu_0},X_{H_{\mu_0},F_{\mu_i}},(S_H)_{\mu_0},(\Delta_H)_{\mu_0})\, .$$
\end{remark}

\begin{remark}\label{remark_newtonian_impact}
  {The results we have presented are valid for arbitrary impact maps. However, in the case of a mechanical system with collisions, the impact map is usually obtained from the Newtonian impact equation (see \cite{brogliato_nonsmooth_1996} for instance).}
  {Let $h\colon Q \to \R$ be a smooth function on $Q$ such that $h^{-1}(0)$ is a submanifold of $Q$. This function will represent a holonomic one-sided contraint on the system, such as a wall. Refer to \cite{ibort_mechanical_1997,ibort_geometric_1998,ibort_geometric_2001} for more details on this type of constraints.}
  The impact map is given by $\Delta_H(q,p)=(q,P_q(p))$, where $P_q:\T_{q}^{*}Q\rightarrow \T_q^{*}Q$ is given by
    \begin{equation}
        P_q(p)=p-(1+e)\frac{\langle\langle p,\dd h_q\rangle\rangle_q}{||\dd h_q||_q^{2}} \dd h_{q},
        \label{impact_newtonian_hamiltonian}
    \end{equation}
     with $||\cdot||_{q}$ denoting the corresponding norm on $\T_{q}^{*}Q$, and $\langle\langle\cdot,\cdot\rangle\rangle_{q}$ is the inner-product on the vector space $\T_{q}^{*}Q$ defined through the kinetic energy of the system as $\displaystyle{\langle\langle\alpha,\beta\rangle\rangle_q=\sum_{i,j=1}^{\dim(Q)}\alpha_j\beta_jW_{ij}^{-1}(q)}$, being $W(q)$ the inertia matrix associated with the mechanical system under study. The parameter $0\leq e\leq 1$ is the coefficient of restitution (for instance, $e=1$ corresponds with elastic impacts and $e=0$ with inelastic impacts). The switching surface is also defined through the inner product as
     $$S_{H}=\{(q,p)\in \cT Q: h(q)=0 \hbox{ and }\langle\langle p,\dd h_q\rangle\rangle_q< 0\}.$$ 
     {Here the condition $\langle\langle p,\dd h_q\rangle\rangle_q< 0$ is simply the requirement that there has to be an strictly positive component of the momenta normal to the switching surface for an impact to occur.}
     Note that the analytical expressions for the switching surface and the impact map depend on the chosen metric. Hence, by choosing different metrics one can obtain different expressions for the impact map and switching surface, which could help to obtain invariant expressions for $S_H$ and $\Delta_H$ for a given action.

     {In this case, a momentum map $J$ is an hybrid momentum map if and only if 
     \begin{equation}
         J\left(q, p-(1+e)\frac{\langle\langle p,\dd h_q\rangle\rangle_q}{||\dd h_q||_q^{2}} \dd h_{q} \right)=\mu_+,
     \end{equation}
     for every $(q, p)\in \cT Q$ such that $h(q)=0,\ \langle\langle p,\dd h_q\rangle\rangle_q< 0$ and $ J(q,p)=\mu_-$.
     }


     Similarly, in a simple hybrid Lagrangian system with a mechanical Lagrangian function $L=\dot q^T W(q) \dot q - V(q)$, the impact can be obtained from the  Newtonian impact equation $P:\T Q\rightarrow \T Q$ given by
    \begin{equation}
      P(q,\dot{q})=\dot{q}-(1+e)\frac{\dd h_{q}\dot{q}}{\dd h_{q}W(q)^{-1}\dd h_{q}^T}W(q)^{-1}\dd h_{q}^T,
      \label{impact_newtonian_lagrangian}
    \end{equation}
     where $W(q)$ is the inertial matrix for the Lagrangian system, $h$ a function describing the switching surface as a submanifold of $Q$ and $e$ the coefficient of restitution.
     The switching surface is $S_L=\left\{(q, \dot q) \colon \T Q\mid h(q)=0 \text{ and } \dd h_q\ \dot q < 0 \right\}$.

    {In the case of a purely kinetical Hamiltonian (or Lagrangian) system, $H=||p||_{q}^2/2$, a Lie group action on $\cT Q$ by isometries preserves the Hamiltonian, and the Newtonian impact map $\Delta_H$ is automatically equivariant.
    }
\end{remark}

\begin{example}[Rolling disk with dissipation hitting fixed walls]\label{example_disk}

\begin{figure}
    \centering
    \begin{tikzpicture}    
        \draw[-,blue,ultra thick] (-3,-2)node[left] {\(y=0\)} -- (5,-2) ;
    
        \draw[-,red,ultra thick] (-3,2)node[left] {\(y=h\)} -- (5,2) ;
    
        \draw[->] (-2.5,-2) -- (3,-2) node[right] {\(x\)};
    
        \draw[->] (-2,-2.5) -- (-2,3) node[above] {\(y\)};
    
        \draw (0,0) circle (1);
    
        \draw[dashed] (0,-1) -- (0,1) node[above] {\(\)};
    
        \draw[-] (0,0) -- ({sin(-40)}, {cos(-40)}) node[midway, above left] {\(\)};
    
        \fill (0,0) circle (1.5pt) node[below left] {\(C\)};
    
        \fill ({sin(-40)}, {cos(-40)}) circle (1.5pt) node[above] {\(P\)};
    
        \draw[thick] (0,0.55) arc[start angle=90, end angle=130, radius=0.55];
    
        \node at (-0.2,0.7) {\(\varphi\)};

        \draw[<->] (0,0)--node[above]{\(R\)}(1,0);
    \end{tikzpicture}
    \caption{Rolling disk with dissipation hitting fixed walls from Example~\ref{example_disk}. Here $C$ is the center of the disk, and $P$ a reference point to measure the angle $\varphi$ from the $Oy$ axis.}
    \label{fig:example_disk}
\end{figure}

Consider a homogeneous circular disk of radius $R$ and mass $m$ moving in the vertical plane $xOy$ (see {\cite[Example 8.2]{ibort_mechanical_1997}}, and also {\cite[Example 3.7]{ibort_geometric_2001}}). {The system is depicted in Figure~\ref{fig:example_disk}.}
Let $(x, y)$ be the coordinates of the centre of the disk and $\varphi$ the angle between a point of the disk and the axis $Oy$. The dynamics of the system is determined by the Hamiltonian $H$ on $\cT (\R^2 \times \mathbb{S}^1)$ given by
\begin{equation}
   H = \frac{1}{2m} \left(p_x^2 +  p_y^2\right) +\frac{1}{2mk^2} p_\varphi^2,
 \end{equation} 
 which is hyperregular since it is mechanical.
The system is subject to external forces given by $F=F_{x}\dd x+F_{y}\dd y$ where $F_{x}=-\frac{2c}{m}(p_{x}xy-p_{y}x^2)$, $F_{y}=\frac{2c}{m}(p_{y}xy-p_{x}y^2)$, for some constant $c>0$. Note that $F_{\varphi}=0$. Forced Hamilton equations of motion are 
\begin{align*}\label{eq:ex1}
\dot p_x&=\frac{2c}{m}(p_{x}xy-p_{y}x^2)\, ,\quad  \dot p_y=-\frac{2c}{m}(p_{y}xy-p_{x}y^2)\,,\quad \dot{p}_{\varphi}=0\, ,\\
\dot{x}&=\frac{1}{m}p_x\,,\quad\dot{y}=\frac{1}{m}p_y\, ,\quad\dot{\varphi}=\frac{1}{k^2m}p_{\varphi}\, .
\end{align*}

Using the Legendre transformation we can obtain the Lagrangian and external force $F^L$. The Lagrangian function $L:\T( \mathbb{R}^{2}\times\mathbb{S}^1)\to\mathbb{R}$ is given by 
$$\displaystyle{
L(x,y,\varphi, \dot{x},\dot{y}, \dot{\varphi})=\left[\frac{1}{2}m(\dot x^2+\dot y^2+k^2\varphi^2)\right]}$$ and $F^L(x,y,\varphi,\dot{x},\dot{y},\dot{\varphi})=F^L_{x}\dd x+F^L_{y}\dd y$ is an external force given by $F^L_{x}=-2c(\dot{x}xy-\dot{y}x^2)$, $F^L_{y}=2c(\dot{y}xy-\dot{x}y^2).$ The forced Euler--Lagrange equations for the free motion of the disk are
\begin{equation*}\label{eq:ex1}
m\ddot x=-2c(\dot{y}x^2-\dot{x}xy), \quad  m\ddot y=2c(\dot{x}y^2-\dot{y}xy),\ddot\varphi=0.
\end{equation*}

Consider the Lie group action of~$\mathbb{S}^1 \times \mathbb{S}^1$ on $Q$ given by $(x, y, \varphi)\mapsto (\cos \alpha\ x- \sin \alpha\ y, \sin \alpha\ x + \cos \alpha\ y, \varphi + \beta)$.
Note that $L$ and $F^L$ are invariant under the lifted action on $\T Q$. The corresponding momentum map is $(J_L)_F(x, y, \varphi, \dot x, \dot y, \dot \varphi)=(mx \dot y- my \dot x, mk^2 \dot \varphi)$.

By introducing polar coordinates $L$ and $F^L$ become
\begin{align*}
L(\theta,r,\varphi,\dot{\theta},\dot{r}, \dot{\varphi})=&\frac{m}{2}(\dot r^2+ r^2\dot \theta^2+k^2\dot\varphi^2),\\ F^L(\theta,r,\dot{\theta},\dot{r})=&2cr^3\dot{\theta}dr,\end{align*} respectively. The forced Euler--Lagrange equations (in polar coordinates) are $$\displaystyle{\ddot r=\left(r-\frac{2c}{m}r^3\right)\dot{\theta},\,\,m\frac{\dd (r^2\dot{\theta})}{\dd t}=0},\,\, mk^2 \ddot{\varphi}=0.$$

The momentum map is now written $(J_L)_F(r,\dot r,\theta,\dot \theta,\varphi,\dot\varphi)= (mr^2\dot\theta,mk^2\dot\varphi)$. By observing the forced Euler--Lagrange equations in polar coordinates, it is clear that $(J_L)_F$ is preserved. Considering $\mu_1=mr^2\dot{\theta}$ and $\mu_2=mk^2\dot{\varphi}$ (i.e., $\dot{\theta}=\frac{\mu_1}{mr^2}$, and  $\dot{\varphi}=\frac{\mu_2}{mk^2}$) the Routhian and the reduced force take the form
\[
R_F^{\mu}(r,\dot{r})=\frac{m}{2}\dot r^2-\frac{\mu_1^2}{2mr^2}-\frac{\mu_2^2}{2mk^2}, \,\, F^L_{\mu}=2cr\frac{\mu_1} {m}\dd r,
\]
and the reduced forced Euler--Lagrange equations for the Routhian $R_F^{\mu}$ and reduced external force $F_{\mu}^{L}$ are given by
\begin{equation*}
\ddot r=\frac{\mu_1^2}{m^2r^3}-2cr\frac{\mu_1} {m^2}.
\end{equation*}

Suppose that there are two rough walls at the axis $y=0$ and at $y=h$, where $h=2\alpha R$ for some constant $\alpha>1$. Assume that the impact with a wall is such that the disk rolls without sliding and that the change of the velocity along the $y$-direction is characterized by an elastic constant $e$. When the disk hits one of the walls, the impact map is given by {(see {\cite[Example 8.2]{ibort_mechanical_1997}}, and also {\cite[Example 3.7]{ibort_geometric_2001}})}
\begin{equation}
  \left(\dot x^-, \dot y^-, \dot \varphi^-  \right)
  \mapsto \left(\frac{R^2 \dot x^- +k^2 R \dot\varphi^-}{k^2+R^2}, - e\dot y^-, \frac{R \dot x^- +k^2 \dot\varphi^-}{k^2+R^2}  \right)\, ,
\end{equation} where {the switching surface is given by $S=C_1\cup C_2$, whose connected components are
\begin{equation}
        \begin{aligned}
                & C_1 = \{(x,y,\varphi,\dot x,\dot y,\dot \varphi)\mid y=R,\,  \dot x =R\dot {\varphi} \hbox{ and } \dot y<0\}\, ,\\
                & C_2 = \{(x,y,\vartheta,p_x,p_y,p_\vartheta)\mid y=h-R,\,  \dot x =R\dot {\varphi} \hbox{ and } \dot y>0\}\, .
        \end{aligned}
\end{equation}
}

For the sake of simplicity, let us assume that $e=1$. It is worth noting that, despite the fact this corresponds to an elastic collision, the momentum map will not be preserved in the impact.



 One can check that $(J_L)_F$ is a generalized hybrid momentum map but not a hybrid momentum map, i.e., $(J_L)_F(q_1, \dot q_1^-)=(J_L)_F(q_2, \dot q_2^-)$ implies that $(J_L)_F(q_1, \dot q_1^+)=(J_L)_F(q_2, \dot q_2^+)$  but  $(J_L)_F(q_1, \dot q_1^+)\neq (J_L)_F(q_1, \dot q_1^-)$ .
Indeed, if $(x,y,\varphi,\dot x^-,\dot y^-, \dot\varphi^-) \in S$, then    
\begin{align*}
    (J_L)_F \circ \Delta (x,y,\varphi,\dot x^-,\dot y^-,\dot\varphi^-)
    &= \left(\hspace{-.1cm}\frac{-m y \left(\hspace{-.05cm}k^2 R \dot{\varphi}^-+R^2 \dot{x}^-\hspace{-.05cm}\right)}{k^2+R^2}-m \dot{y}^- x, \frac{k^2 m \left(\hspace{-.05cm}k^2 \dot{\varphi}^-+R \dot{x}^-\hspace{-.05cm}\right)}{k^2+R^2}\hspace{-.1cm}\right)\\
    &= \left(-mx\dot y^{-} -mR^2 \dot{\varphi}^{-}, mk^2 R \dot \varphi^{-}\right),
\end{align*}
where in the last step we have used that $y=R$ and $\dot x^{-}=R\dot \varphi^{-}$ (for the wall at $y=h$ the result is analogous).

In polar coordinates, for $\theta=\arctan (y/x)$, we have 
\begin{equation}
\begin{aligned}
\dot \theta^+&=\frac{1}{1+(y/x)^2} \left(\frac{\dot y^+ x-y \dot x^+ }{x^2}\right)
=\frac{1}{r^2} \left(-\dot y^- x - y\frac{R^2 \dot x^-+k^2 R \dot \theta^-}{R^2+k^2}\right)
=\frac{1}{r^2} \left(-\dot y^- x - y\dot x \right)
=-\dot \theta^-
\label{impact_theta_polar}
\end{aligned}
\end{equation}
where we have replaced the expression for $\dot x^+$, $\dot y^+$ and we have used that $x^2+y^2=r^2$, $\left(y \dot x^- -\dot y^- x\right)=-r^2\dot\theta^-$ and $\dot x^-=R\dot \theta^-$. Moreover,
\begin{equation}
\begin{aligned}
   \dot r^+ &= \frac{1}{r} (x \dot x^+ + y \dot y^+)
   = \frac{1}{r} (x \dot x^- - y \dot y^-)
   = (2\cos^2 \theta -1) \dot r^- -2r \sin \theta \cos \theta \dot \theta^-,
  \label{reduced_impact_map}
\end{aligned}
\end{equation}
and
\begin{equation}
\begin{aligned}
\dot\varphi^+&=\frac{R \dot x^- +k^2 \dot\varphi^-}{k^2+R^2} 
= \frac{R \left(\cos \varphi\, \dot r^-  -\sin \varphi\, \dot \varphi^- \right) +k^2 \dot\varphi^-}{k^2+R^2}.
\label{impact_vartheta_polar}
\end{aligned}
\end{equation}
{The connected components of the switching surface can be written as
\begin{equation}
\begin{aligned}
        C_1 & = \left\{(r,\theta,\varphi,\dot r,\dot \theta,\dot \varphi)\mid r\sin \theta=R\, , \quad 
        \dot r \cos \theta - r \dot \theta \sin \theta=R\dot {\varphi}
        \quad
        \hbox{and} 
        \quad 
        \dot r \sin \theta + r \dot \theta \cos \theta < 0
        \right\}\, , \\
        C_2 & = \left\{(r,\theta,\varphi,\dot r,\dot \theta,\dot \varphi)\mid r\sin \theta=h-R\, , \quad 
        \dot r \cos \theta - r \dot \theta \sin \theta=R\dot {\varphi}
        \quad
        \hbox{and} 
        \quad 
        \dot r \sin \theta + r \dot \theta \cos \theta > 0
        \right\}\, .
\end{aligned}
\end{equation}
}
Let $(\mu_1^-,\mu_2^-)$ and $(\mu_1^+,\mu_2^+)$ be the value of the momentum map before and after an impact, respectively. We can write $\dot\theta^\pm=\mu_1^\pm/mr^2$ and $\dot \varphi^\pm=\mu_2^\pm/mk^2$, so the reduced switching map is
$$\Delta_{\mu_1, \mu_2}\colon \dot r^- \mapsto  (2\cos^2 \theta -1) \dot r^- -2r \sin \theta \cos \theta \frac{\mu_1^-}{mr^2}\, ,$$
with the relations
$$\mu_1^+=-\mu_1^-,\quad \mu_2^+=\mu_2^-,$$
which are derived from Eqs.~\eqref{impact_theta_polar} and \eqref{impact_vartheta_polar}, respectively. 
{Note that in this example it is straightforward to calculate the value of the momentum map after an arbitrary number of impacts, simplifying the reconstruction procedure. Indeed, if the initial value (after the ``$0$-th impact'') of the momentum map is $\mu_0 = (\mu_{0,1}, \mu_{0,2})$, its value after the $j$-th impact will be $\mu_i=\Big((-1)^{j} \mu_{0,1}, \mu_{0,2}\Big)$. Therefore, after the $j$-th impact, the Routhian and the reduced force take the form
$$R_F^{\mu_{j}} = R_F^{\mu_0} = \frac{m}{2}\dot r^2-\frac{\mu_{0,1}^2}{2mr^2}-\frac{\mu_{0,2}^2}{2mk^2}\, , \quad 
F^L_{\mu_{j}} =(-1)^{j} F^L_{\mu_0} =  (-1)^{j}\, 2cr\frac{\mu_{0,1}} {m}\dd r\, ,$$
with forced Euler--Lagrange equations
\begin{equation*}
\ddot r=\frac{\mu_{0,1}^2}{m^2r^3} - (-1)^{j}\, 2cr\frac{\mu_{0,1}} {m^2}\, .
\end{equation*}
The reduced switching map and the connected components of the reduced switching surface after the $j$-th impact are given by
\begin{align*}
    \Delta_{\mu_{j}} (\dot r^-) & = (2\cos^2 \theta -1) \dot r^- - (-1)^j\, 2r \sin \theta \cos \theta \frac{\mu_{0,1}}{mr^2}\, ,\\
    C_{1,\mu_{j}}&=\Big\{(r,\dot r)\mid r\sin \gamma=R\, , \
    \dot r \cos \gamma - \frac{(-1)^j\, \mu_{0,1}}{mr} \sin \gamma=R\frac{\mu_{0,2}}{mk^2}\, ,
    \\ &  \quad
    \hbox{ and }
    \dot r \sin \gamma + \frac{(-1)^j\, \mu_{0,1}}{mr} \cos \gamma < 0
    \hbox{ for some } \gamma \in [0, 2\pi)\Big\}\, , \\
    C_{2,\mu_{j}}&=\Big\{(r,\dot r)\mid r\sin \gamma=h-R\, , \
    \dot r \cos \gamma -  \frac{(-1)^j\, \mu_{0,1}}{m} \sin \gamma=R\frac{\mu_{0,2}}{mk^2}\, ,
    \\ &  \quad
    \hbox{ and }
    \dot r \sin \gamma + \frac{(-1)^j\, \mu_{0,1}}{mr} \cos \gamma > 0
    \hbox{ for some } \gamma \in [0, 2\pi)\Big\}\, ,
\end{align*}
In particular, observe that these reduced objects only depend on whether the number of the last impact was even or odd.
}
\end{example}

\begin{example}[Non-abelian reduction: Wong's equations]
    Consider a Riemannian manifold $(Q, \mathcal{G})$ on which a Lie group $G$ acts freely and properly by isometries, and such that the restriction of the metric $\mathcal{G}$ to the fibers of $Q\to Q/G$ comes from a bi-invariant metric on $G$ (see \cite{crampin_rouths_2008}). 
    Let $(q^i, q^a)$ be fibered coordinates on $Q\to Q/G$, and $(q^i, q^a, \dot q^i, \dot q^a)$ be the induced coordinates on $\T Q$.
    The geodesic equations for $\mathcal G$ can be derived from the Lagrangian function
    \begin{equation}
        L = \frac{1}{2} g_{ij} \dot q^i \dot q^j 
        + \frac{1}{2} g_{ab} \dot q^a \dot q^b\, ,
    \end{equation}
    where $\mathcal{G} = g_{ij} \dd q^i \otimes \dd q^j + g_{ab} \dd q^a \otimes \dd q^b$.
    The momentum map of the lifted $G$-action on $\T Q$ is given by 
    \begin{equation}
        J(q^i, q^a, \dot q^i, \dot q^a) = (g_{ab} \dot q^b).
    \end{equation}
    The Routhian is
    \begin{equation}
        R^\mu = \frac{1}{2} g_{ij} \dot q^i \dot q^j 
        - \frac{1}{2} g^{ab} \mu_a \mu_b,
    \end{equation}
    where $(g^{ab})$ is the inverse matrix of $(g_{ab})$, and $\mu=(\mu_a)$ is a regular value of $J$. The reduced geodesic equations are known as Wong's equations.


   Let $h$ be a function on $Q$, defining the Newtonian impact map $\Delta_L$ and the switching surface $S_L$ (see Remark \ref{remark_newtonian_impact}). 
   Suppose that $h$ is the lift of a funtion on $Q/G$, namely, $h=h(q^i)$. Then,
    \begin{equation}
       \Delta_L \left( q^i, q^a, \dot q^ i, \dot q^a \right)
        = \left( q^i, q^a, \dot{q}^i-(1+e)\frac{\dd h_{q}\dot{q}}{\dd h_{q}\mathcal G^{-1}\dd h_{q}^T}g^{ij} \frac{\partial h}{\partial q^j}
        , \dot q^a \right),
    \end{equation}
    and
    \begin{equation}
       S_L = \left\{( q^i, q^a, \dot q^ i, \dot q^a ) \in \T Q \mid h(q^i)=0 \text{ and } \frac{\partial h}{\partial q^i}\dot q^i < 0 \right\}.   
    \end{equation}
    Since $J$ only depends on $(q^a, \dot q^a)$ and $\Delta_L$ does not change these coordinates, $J(x)=\mu_-$ for any $x\in \Delta\left(J_{\vert S_L}^{-1}(\mu_-)\right)$, i.e., $J$ is a hybrid momentum map.
\end{example}

\section{Extension to the reduction for time-dependent forced mechanical systems with symmetries}\label{section_cosymp_reduction}
 
Next we extend the reduction of simple hybrid forced Hamiltonian systems to the case in which the Hamiltonian, the external force and the switching surface depend explicitly on time. The geometric framework that we will employ for non-autonomous mechanics will be a cosymplectic manifold \cite{cantrijn_gradient_1992}. In particular, in this section we extend the results in \cite{colombo_note_2020} to forced mechanical systems. 

Consider a time-dependent forced Lagrangian $L:\mathbb{R}\times \T Q\to\mathbb{R}$ with a time-dependent external force denoted by $F^{L}$, a semibasic 1-form on {$\mathbb{R}\times \T Q\to Q$}. Let us denote by $\mathbb{F} L\colon \mathbb{R}\times \T Q\to \mathbb{R}\times \cT Q$ the Legendre transformation associated with $L$, i.e., the map $(t,q,\dot q)\mapsto (t,q,p=\partial L/\partial \dot q)$. Hereinafter, assume that the Lagrangian is hyperregular,\footnote{In particular, this always holds for mechanical Lagrangians.}, i.e., that $\mathbb{F} L$ is a diffeomorphism. This permits to work out the velocities $\dot q$ in terms of $(t,q,p)$ by means of the inverse of $\mathbb{F} L$, and to define the Hamiltonian function  $H\colon \mathbb{R}\times \cT Q\to \mathbb{R}$ as $H(t,q,p)=\langle p, \dot q(t,q,p)\rangle - L(t, q,\dot q(t,q,p))
$, and the external force $F$ such that $\mathbb{F} L(F^L)=F$, a semibasic 1-form on $\mathbb{R}\times \cT Q.$

In order to characterize the dynamics of a non-autonomous forced Hamiltonian system, consider the manifold $\mathbb{R}\times \cT Q$ equipped with the canonical cosymplectic structure $\omega=\dd q^i\wedge \dd p_i,\eta=\dd t$, where $(q^i)$ are local coordinates on $Q$, and $(t, q^i, p_i)$ are the induced coordinates on $\mathbb{R}\times \cT Q$.
There is an unique vector field $\Reeb$ on $\mathbb{R}\times \cT Q$ such that $\eta (\Reeb)=1$ and $\contr{\Reeb}\omega=0$, called the \myemph{Reeb vector field} which in coordinates reads, $\Reeb =\partial_t$.
 The cosymplectic structure defines an isomorphism between vector fields and 1-forms on $\R \times \cT Q$. In addition, with every smooth function $f$ on $\R\times \cT Q$, one can associate a \myemph{Hamiltonian vector field} $X_f$, such that $\contr{X_f} \eta = 0$ and $\contr{X_f}\omega = \dd f - \Reeb (f) \eta$. We can also define the \myemph{evolution vector field} $X_{f,t}=X_f + \Reeb $.

Given a time-dependent Hamiltonian function $H(t,q,p )$ and a time-dependent external force $F(t,q,p)$, the \myemph{forced Hamiltonian vector field} $X_{H,F}$ is the vector field on $\mathbb{R}\times \cT Q$ defined by $X_{H,F}=X_{H}+Z_F$, where \begin{align*}\contr {X_{H}}\omega_Q&=\dd H-\Reeb (H)\eta,\, \contr {X_H}\eta=0,\\ \contr {Z_{F}}\omega_Q&=F-F(\Reeb) \eta,\quad \contr {Z_F}\eta=0.\end{align*}
The \myemph{forced evolution vector field} corresponding to the forced Hamiltonian system $(H, F)$, denoted by $X_{H,F,t}$, is given by $$X_{H,F,t}=\Reeb +X_{H,F}=\frac{\partial}{\partial t}+\frac{\partial H}{\partial p_i}\frac{\partial}{\partial q^i}-\left(\frac{\partial H}{\partial q^i}+F_{i}\right)\frac{\partial}{\partial p_i}.$$

A similar procedure can be used in the Lagrangian framework. A hyperregular Lagrangian defines a cosymplectic structure on $\mathbb{R}\times \T Q$, given by $\omega_L=\mathbb{F} L^*(\dd q_i\wedge \dd p_i)=\dd q^i\wedge \dd\left(\frac{\partial L}{\partial\dot{q}^i}\right), \eta=\dd t$ (with a slight abuse of notation, the same symbol $\eta=\dd t$ will be used for two different 1-forms on different manifolds). Let $\Reeb_L$ denote the associated Reeb vector field. 

 The \myemph{{Lagrangian} energy} $E_L\colon \mathbb{R}\times \T Q\to \mathbb{R}$ is given by $E_L(t,q,\dot q)=\langle \mathbb{F} L(t,q,\dot q),\dot q\rangle - L(t,q,\dot q),$ from which one can compute the Hamiltonian forced vector field $X_{L, F^L}$ associated with $E_L$ and $F^L$ via the Lagrangian cosymplectic structure. This leads to a forced evolution vector field $X_{L,F^{L},t}=X_{L,F^{L}}+\Reeb_L$. 
Finally, the equivalence between the Lagrangian and Hamiltonian dynamics in the hyperregular case for forced time-dependent systems is achieved via $\mathbb{F} L$ as follows.


\begin{proposition}\label{Proposition12} 
The tangent map of $\F L$ maps $X_{L,F^{L},t}$ onto $X_{H,F,t}$. In other words $(\T\F L) X_{L,F^{L},t}=X_{H,F,t}$, where $(\T\F L)\colon \T( \T Q)\to \T( \cT Q).$ In particular, the flow of $X_{L,F^{L},t}$ is mapped onto the flow of $X_{H,F,t}$. 
\end{proposition}

\begin{proof} The evolution vector field $X_{H,F,t}$ is characterized by $\contr {X_{H}}\omega_Q=\dd H-\Reeb (H)\eta,\, \contr {X_H}\eta=0,\, \contr {Z_{F}}\omega_Q=F-\Reeb (F)\eta\hbox{ and } \contr {Z_F}\eta=0.$ Observe that 
\begin{align*}
(\mathbb{F} L)^{*}(\contr{X_{H}}\omega_{Q})
&=(\mathbb{F} L)^{*}(\dd H-\Reeb (H)\eta)
=(\mathbb{F} L)^{*}(\dd H)-(\mathbb{F} L)^{*}(\Reeb (H)\eta)
=\dd ((\mathbb{F} L)^{*}H)-\Reeb ((\mathbb{F} L)^{*}H)\eta
\\&
=\dd (E_{L})-\Reeb (E_{L})\eta
=\contr{X_{L}}\omega_{L}.
\end{align*}
This means that 
\begin{align*}\contr {X_{L}}\omega_{L}&=(\mathbb{F} L)^{*}(\contr{X_{H}}\omega_{Q})=\contr{(\mathbb{F} L^{-1})_{*}X_{H}}(\mathbb{F} L^{*}\omega_{Q})
=\contr{(\mathbb{F} L^{-1})_{*}X_{H}}\omega_{L}.\end{align*} 
Similarly, one can show that $\contr {X_L}\eta=\contr{(\mathbb{F} L^{-1})_{*}X_{H}}\eta$. This implies that $X_{L}=(\mathbb{F} L^{-1})_{*}X_{H}$, that is, $(\F L)_{*} X_{L}=X_{H}$.

On the other hand, observe that 
\begin{align*}
(\mathbb{F} L)^{*}(\contr{Z_{F}}\omega_{Q})
&=(\mathbb{F} L)^{*}(F-\Reeb (F)\eta)
=(\mathbb{F} L)^{*}(F)-(\mathbb{F} L)^{*}(\Reeb (F)\eta)
\\&
=(\mathbb{F} L)^{*}F-\Reeb ((\mathbb{F} L)^{*}F)\eta=F^L-\Reeb (F^{L})\eta
=\contr{Z_{F^L}}\omega_{L},
\end{align*}
which implies that
 \begin{align*}\contr {Z_{F^L}}\omega_{L}&=(\mathbb{F} L)^{*}(\contr{Z_{F}}\omega_{Q})=\contr{(\mathbb{F} L^{-1})_{*}Z_{F}}(\mathbb{F} L^{*}\omega_{Q})
 =\contr{(\mathbb{F} L^{-1})_{*}Z_{F}}\omega_{L}.\end{align*} 
 Similarly, one can see that $\contr {Z_{F^L}}\eta=\contr{(\mathbb{F} L^{-1})_{*}Z_{F}}\eta$. Hence, $Z_{F^L}=(\mathbb{F} L^{-1})_{*}Z_{F}$, that is, $(\F L)_{*} Z_{F^L}=Z_{F}$. \end{proof}

\begin{definition} A \myemph{simple hybrid time-dependent forced Lagrangian system} is described by the tuple $\mathscr{L}^t_{F}=(\R\times \T Q,X_{L,F^{L},t},$ $S^t_{L},\Delta^t_{L})$, where
 $Q$ is a differentiable manifold, $X_{L,F^{L},t}$ is the forced evolution vector field associated with the time-dependent forced Lagrangian system $(L, F^L)$, $S^t_{L}$ is an embedded submanifold of $\mathbb{R}\times \T Q$ with co-dimension one, the {switching surface}, and $\Delta^t_{L}\colon S^t_{L}\to \mathbb{R}\times \T Q$ is a smooth embedding, the {impact map}.

 {For physical reasons, we will assume that the impact map fixes time, i.e., $\Delta^t_{L}\colon S^t_{L}\to \mathbb{R}\times \T Q$ projects to an embedding $\overline{\Delta^t_{L}}\colon \pi_{\T Q}({S^t_{L}})\to \T Q$, with $\pi_{\T Q}\colon \T Q \times \R \to \T Q$ the canonical projection.}
\end{definition}
Analogously, one can introduce the notion of \myemph{simple hybrid time-dependent forced Hamiltonian system} $\mathscr{H}^t_{F}=(\R\times \cT Q,X_{H,F,t},$ $S^t_H,\Delta^t_H)$, where $X_{H,F,t}$ is is the forced evolution vector field associated with the time-dependent forced Hamiltonian system $(H, F)$. The relation between both hybrid flows is given by the following result, based on the equivalence between the Lagrangian and Hamiltonian dynamics in the hyperregular case can achieved via the fiber derivative $\mathbb{F}H$.

\begin{proposition}\label{Proposition5}
If $\chi^{\mathscr{H}^t_{F}}=(\Lambda,\mathcal{J},\mathscr{C})$ is a hybrid flow for $\mathscr{H}^t_{F}$, $S^t_L=\F H(S^t_H)$, and $\Delta^t_L$ is defined in such a way that $\F H\circ \Delta^t_H=\Delta^t_L\circ \restr{\F H}{S_H}$, then $\chi^{\mathscr{L}_F}=(\Lambda,\mathcal{J},(\F H)(\mathscr{C}))$ with $(\F H)(\mathscr{C})=\{(\F H)(c_{i})\}_{i\in \Lambda}$.
\end{proposition}%

\begin{proof} The proof follows straightforwardly from the Proof of Proposition \ref{Proposition2}.\end{proof}%

\vspace{.4em}

Let $\mathscr{L}^t_{F}=(\R\times \T Q,X_{L,F^{L},t},S^t_{L},\Delta^t_{L})$ be a simple hybrid time-dependent forced Lagrangian system and let $\psi\colon G\times Q\to Q$ be a free and proper Lie group action with $\psi^{\R\times \cT Q}$ denoting its natural lift, namely, $G$ acts on $\R$ by the identity and on $\cT Q$ by $\psi^{\cT Q}$. The action $\psi^{\R\times \cT Q}$ enjoys the following properties \cite{abraham_foundations_2008,de_leon_symmetries_2021}:
\begin{itemize}
	\item It is a cosymplectic action, meaning that $(\psi^{\R\times \cT Q}_g)^*\omega=\omega$ and $(\psi^{\R\times \cT Q}_g)^*\eta=\eta$ for every $g\in G$
	\item It admits an $\hbox{Ad}^*$-equivariant momentum map 
	$
	\bar{J}\colon \R\times \cT Q\to \mathfrak{g}^*$ given by
	$\langle \bar{J}(t,q,p), \xi\rangle=\langle p, \xi_Q\rangle$, for each $\xi\in\mathfrak{g}$, where $\xi_Q(q)=\dd (\psi_{\exp(t\xi)}q)/dt$ is the infinitesimal generator of  $\xi\in \mathfrak{g}.$
\end{itemize}
Likewise, $\psi^{\R\times \T Q}$ denotes the natural lift action of $G$ on $\R\times \T Q$, i.e., $G$ acts on $\R$ by the identity and on $\T Q$ by $\psi^{\T Q}$.

Let $X$ be a vector field on $Q$. The complete lift of $X$ at $(t, v_q)\in \R\times \cT Q$ is given by $X^c_{t, v_q}=(0_t, \tilde X^c(v_q))$, where
$\tilde X^c(v_q)$ denotes the complete lift of $X$ to $\cT Q$ at $v_q$. Locally, $X^c=X^i\frac{\partial}{\partial q^i}-p_j\frac{\partial X^j}{\partial q^i}\frac{\partial}{\partial p_i}$. Analogously, one can define the complete lift of $X$ to $\R\times \T Q$, locally, $X^c=X^i\frac{\partial}{\partial q^i}+\dot{q}^j\frac{\partial X^i}{\partial q^j}\frac{\partial}{\partial \dot{q}^i}$.

As in the autonomous case, a function $f$ on $\R\times \cT Q$ is called a \myemph{constant of the motion} (or a \myemph{conserved quantity}) for $(H, F)$ if it takes a constant value along the trajectories of the system or, in other words, $X_{H,F,t}(f)=0$.

For each $\xi\in\mathfrak{g}$ and $\alpha_{q}\in \cT Q$, consider the function $\bar{J}^{\xi}:\R\times \cT Q\rightarrow\R$ given by $\bar{J}^{\xi}(t,\alpha_{q})=\langle \bar{J}(t,\alpha_{q}),\xi\rangle$. If $H$ and $F$ are $G$-invariant, then $\bar{J}^{\xi}$ is a conserved quantity for $(H,F)$ if and only if $F(\xi_{Q}^{c})=0.$ In addition, the vector subspace of $\mathfrak{g}$ given by $\mathfrak{g}_{F}=\{\xi\in \mathfrak{g}:F(\xi_{Q}^{c})=0,\,\, \contr {\xi_{Q}^{c}}\dd F=0\}$ is a Lie subalgebra of $\mathfrak{g}$.  

As in the autonomous case, to perform a hybrid reduction one needs to impose some compatibility conditions between the action and the hybrid system (see \cite{ames_hybrid_2006-1} and \cite{ames_hybrid_2006}). By a \myemph{hybrid action} on the simple hybrid system $\mathscr{H}^t_{F}=(\R\times \cT Q,X_{H,F,t},$ $S^t_H,\Delta^t_H)$ we mean a Lie group  action $\psi\colon G\times Q\to Q$ such that
\begin{itemize}
  \item $H$ is invariant under $\psi^{\R\times \cT Q}$, i.e. $H\circ \psi^{\R\times \cT Q}=H$.
  \item $\psi^{\R\times \cT Q}$ restricts to an action of $G$ on $S^t_H$.
  \item $\Delta^t_H$ is equivariant with respect to the previous action, namely 
  $$\Delta^t_H\circ \restr{\psi^{\R\times \cT Q}_g}{S^t_H}=\psi^{\R\times \cT Q}_g\circ \Delta^t_H.$$
\end{itemize}

From the Cosymplectic Reduction Theorem \cite{albert_theoreme_1989} (see also \cite{de_leon_cosymplectic_1993}), we can obtain the non-autonomous analogue of Theorem \ref{theorem_reduction_autonomous}.

\begin{theorem} \label{theorem_reduction_non_autonomous}
Let $\mathscr{H}_{F}^t=(\R\times \cT Q,X_{H,F,t}, S_H^t,\Delta_H^t)$ be a time-dependent simple hybrid forced Hamiltonian system. Let $\psi:G\times Q \to Q$ be a hybrid action of a connected Lie group $G$ on $Q$.
Suppose that $H$ and $F$ are $G_F$-invariant and assume that $\bar{J}_F$ is a generalized hybrid momentum map.
Consider a sequence $\left\{\mu_i  \right\}$ of regular values of $\bar{J}_F$, such that $$\Delta_H^t \left(\restr{\bar{J}}{S_H^t}^{-1}(\mu_i)  \right)\subset \bar{J}^{-1}(\mu_{i+1}).$$
Let $(G_{F})_{\mu_i}=(G_{F})_{\mu_0}$ be the isotropy subgroup in $\mu_i$
 under the co-adjoint action. Then,
\begin{enumerate}[(i)]
\item $\bar{J}_{F}^{-1}(\mu_i)$ is a submanifold of $\R\times \cT Q$ and $X_{H,F,t}$ is tangent to it.
\item  The reduced space $M_{\mu_i}\coloneqq \bar{J}_{F}^{-1}(\mu_i)/(G_{F})_{\mu_0}$ is a cosymplectic manifold, whose cosymplectic structure $(\omega_{\mu_i}, \eta_{\mu_i})$ is uniquely determined by  $\pi^{*}_{\mu_i}\omega_{\mu_i}=\incl^{*}_{\mu_i}\omega_{Q}$ and $\pi^{*}_{\mu_i}\eta_{\mu_i}=\incl^{*}_{\mu_i}\eta$, where $\pi_{\mu_i}:J_{F}^{-1}(\mu_i)\rightarrow M_{\mu_i}$ and $\contr {\mu_i}:\bar{J}_{F}^{-1}(\mu_i)\hookrightarrow \R\times \cT Q$ denote the canonical projection and the  canonical inclusion, respectively. In addition, the Reeb vector field $\Reeb$ projects onto $\Reeb_{\mu_i}$, the Reeb vector field defined by $\omega_{\mu_i}$ and $\eta_{\mu_i}$. 
 \item 
 $(H,F)$ induces a reduced time-dependent forced Hamiltonian system $(H_{\mu_i}, F_{\mu_i})$ on $M_{\mu_i}$, given by $H_{\mu_i}\circ\pi_{\mu_i}=H\circ \contr {\mu_i}$ and $\pi^{*}_{\mu_i}F_{\mu_i}=\incl^{*}_{\mu_i}F$. Moreover, the forced evolution vector field $X_{H,F,t}$ projects onto $X_{H_{\mu_i},F_{\mu_i},t}$.
 \item $\restr{\bar{J}_{F}}{S_H^t}^{-1}(\mu_i)\subset S_H^t$ reduces to a submanifold of the reduced space
   $(S_H^t)_{\mu_i}\subset \bar{J}_{F}^{-1}(\mu_i)/(G_{F})_{\mu_0}.$ 
 \item 
 $\Delta_{H\mid \bar{J}^{-1}(\mu_i)}^t$ reduces to a map $(\Delta_H^t)_{\mu_i}:(S_H^t)_{\mu_i} \rightarrow \bar{J}_{F}^{-1}(\mu_{i+1})/(G_{F})_{\mu_0}.$
\end{enumerate}
{Therefore, after the reduction procedure, we get a sequence of reduced time-dependent simple hybrid forced Hamiltonian systems $\left\{\mathscr{H}_ {F}^{\mu_i}  \right\}$, where 
$\mathscr{H}_ {F}^{\mu_i}=(\bar{J}_{F}^{-1}(\mu_i)/(G_{F})_{\mu_0},$ $X_{H_{\mu_i},F_{\mu_i}},(S_H^t)_{\mu_i},(\Delta_H^t)_{\mu_i})$.}

\begin{center}
\begin{tikzcd}
\cdots \arrow[r] & \bar{J}_F^{-1}(\mu_i) \arrow[dd]        &  & \restr{\bar{J}_F}{S_H^t}^{-1}(\mu_i) \arrow[dd] \arrow[rr, "\Delta_{H\mid \bar{J}^{-1}(\mu_i)}"] \arrow[ll, hook] &  & \bar{J}_F^{-1}(\mu_{i+1}) \arrow[dd]        & \cdots \arrow[l, hook] \\
                 & {} \arrow[d]                      &  &                                                                                               &  &                                       &                        \\
\cdots \arrow[r] & \frac{\bar{J}_F^{-1}(\mu_i)}{G_{\mu_0}} &  & \left(S_H^t\right)_{\mu_i} \arrow[rr, "\left(\Delta_H^t\right)_{\mu_i}"] \arrow[ll, hook]         &  & \frac{\bar{J}_F^{-1}(\mu_{i+1})}{G_{\mu_0}} & \cdots \arrow[l, hook]
\end{tikzcd}

\end{center}

\end{theorem}
\begin{proof} The proof follows straightforwardly from the Proof of Proposition \ref{theorem_reduction_autonomous}.\end{proof}

By a \myemph{hybrid action} on the simple hybrid system $\mathscr{L}^t_{F}=(\T Q,X_{L,F^L,t}, S^t_{L},\Delta^t_{L})$ we mean a Lie group  action $\psi\colon G\times Q\to Q$ such that
\begin{itemize}
  \item $L$ is invariant under $\psi^{\R\times \T Q}$, i.e. $L\circ \psi^{\R\times \T Q}=L$.
  \item $\psi^{\R\times \T Q}$ restricts to an action of $G$ on $S^t_L$.
  \item $\Delta^t_L$ is equivariant with respect to the previous action, namely $$\Delta^t_L\circ \restr{\psi^{\R\times \T Q}_g}{S^t_L}=\psi^{\R\times \T Q}_g\circ \Delta^t_L.$$
\end{itemize}


As in the autonomous case, the reduction picture in the Lagrangian side can now be obtained from the Hamiltonian one by adapting the scheme developed in~\ref{section_symp_reduction} to the cosymplectic setting. Suppose that $\mathscr{L}^t_{F}$ is equipped with a hybrid action $\psi$ and that $(L,F^{L})$ is $G_{F^L}$-regular. Since $L$ is invariant and hyperregular, the Legendre transformation $\mathbb{F} L$ is a diffeomorphism such that:
\begin{itemize}
	\item It is equivariant with respect to $\psi^{\R\times \T Q}$ and $\psi^{\R\times \cT Q}$,
	\item It preserves the level sets of the momentum map, that is, $\mathbb{F} L ((\bar{J}_L)_{F}^{-1}({\mu_i}))=\bar{J}_{F}^{-1}({\mu_i}),$
	\item It relates both cosymplectic structures, that is, $(\mathbb{F} L)^*\omega_{Q}=\omega_L$ and $(\mathbb{F} L)^*\eta=\eta$, that is,  $\mathbb{F} L$ is a \myemph{cosymplectomorphism}.
\end{itemize}
The equivariance of $\mathbb{F} L$ implies that $\psi^{\R\times \T Q}$ admits an $\hbox{Ad}^*$-equivariant momentum map $\bar{J}_L: \R\times \T Q\to\mathfrak{g}^{*}$ given by $\bar{J}_L=\bar{J}\circ\mathbb{F} L$. 

It follows that the map $\mathbb{F} L$ reduces to a cosymplectomorphism $(\mathbb{F} L)_{\text{red}}$ between the reduced spaces. Therefore we get the following commutative diagram
$$
\begin{tikzcd}[column sep=1.2cm, row
sep=1.2cm]
(\mathbb{R}\times \T Q,S^t_L,\Delta^t_L)  \arrow[d,swap,"\text{Red.}"] \arrow[r,"\mathbb{F} L"] & (\mathbb{R}\times \cT Q,S^t_H,\Delta^t_H) \arrow[d,"\text{Red.}"]\\
(\bar{\mathcal{M}}^{L}_{\mu_0},(S^t_L)_{\mu_i},(\Delta^t_L)_{\mu_i}) \arrow[r,"(\mathbb{F} L)_{\text{red}}"] & (\bar{\mathcal{M}}^{H}_{\mu_0},(S^t_H){{\mu_i}},(\Delta^t_H)_{{\mu_i}})
\end{tikzcd}
$$
where we have used the notation $\bar{\mathcal{M}}^{L}_{\mu_0}\coloneqq (\bar{J}_L)_{F}^{-1}(\mu_i)/(G_{F^{L}})_{\mu_0}$ and $\bar{\mathcal{M}}^{H}_{\mu_0}\coloneqq \bar{J}_{F}^{-1}(\mu_i)/(G_{F})_{\mu_0}$.

By assuming $G_F$-regularity we obtain the identification 
\begin{equation*}
(\bar{J}_L)_F^{-1}({\mu_i})/(G_{F})_{\mu_0}\simeq \mathbb{R}\times \left(\T( Q/G)\times_{Q/G}Q/(G_{F})_{\mu_0}\right)\, ,
\end{equation*}
{where $\times_{Q/G}$ denotes the fiber product, i.e., the subset of the Cartesian product consisting on pairs $(a, b)$ such that the projections of $a$ and $b$ on $Q/G$ coincide.}
It is possible to interpret the reduced dynamics on this space as being the Lagrangian dynamics of some regular time-dependent Lagrangian subjected to a time-dependent gyroscopic force (arising from the magnetic term) if one works in the class of magnetic Lagrangians~\cite{langerock_routh_2012}, which in the present situation should be extended to include time-dependent Lagrangians and external forces. The Routhian is defined as (the reduction of) the $(G_{F})_{\mu_0}$-invariant function
\begin{equation}\label{eq:Routhiandefinition}
R_F^{\mu_i}=L-\mathcal{A}_{\mu_i}
\end{equation}
restricted to $(\bar{J}_L)_F^{-1}({\mu_i})$. 
Then, the following diagram commutes:
$$
\begin{tikzcd}[column sep=1cm, row
sep=1.2cm]
\mathbb{R}\times \T Q  \arrow[d,swap,"\text{Red.}"] \arrow[r,"\mathbb{F} L"] & \mathbb{R}\times \cT Q \arrow[d,"\text{Red.}"]\\
 \mathbb{R}\times \left(\T( Q/G)\times_{Q/G}Q/(G_{F})_{\mu_0}\right) \arrow[r,"\mathbb{F} R_F^{\mu_i}"] &  \mathbb{R}\times \left(\cT (Q/G)\times_{Q/G}Q/(G_{F})_{\mu_0}\right)
\end{tikzcd} 
$$
\begin{remark}
Assume we work with $Q=\mathbb{S}^1\times M$, where $M$ is called the shape space and the action is  $(\theta,x)\mapsto(\theta+\alpha,x)$. The forced Lagrangian system has a cyclic coordinate $\theta$, i.e., $L$ is a function of the form $L(t,\dot{\theta},x,\dot{x})$,  and $F$ is of the form $F(t,\dot{\theta},x,\dot{x}) = F_{x}(t,\dot{\theta},x,\dot{x}) \dd x$. The conservation of the momentum map $(\bar{J}_L)_{F}={\mu_i}$ reads $\frac{\partial L}{\partial \dot{\theta}}={\mu_i},$ and one can use this relation to express $\dot{\theta}$ as a function of the remaining --non cyclic-- coordinates and their velocities, and the prescribed regular value of the momentum map ${\mu_i}$, i.e., $\dot \theta = \dot \theta(t,x, \dot x, \mu_i)$. Note that this is the stage at which the $G_F$-regularity of $L$ and $F^{L}$ is used: it guarantees that $\dot{\theta}$ can be worked out in terms of $x$, $\dot{x}$ and ${\mu_i}$. If one chooses the canonical flat connection on $Q\rightarrow Q/\mathbb{S}^1=M,$ then the Routhian can be computed as
where the notation means that we have everywhere expressed $\dot{\theta}$ as a function of $(t,x,\dot x,{\mu_i})$. 


As in the non-autonomous case, collisions with the switching surface will, in general, modify the value of the momentum map (non-elastic case).  Therefore, if $\mathcal{J}=\{I_{i}\}_{i\in \Lambda}$ is the hybrid interval (see Definition~\ref{def:flow}), the Routhian has to be defined in each $I_i$ taking into account the value of the momentum $\mu_i$ after the collision at time $\tau_i$. Note that this also has an influence in the way the impact map $\Delta^t_L$ and the switching $S$ are reduced. Let us denote: (1) $\mu_i$ the momentum of the system in $I_i=[\tau_i,\tau_{i+1}]$, (2) $(\Delta^t_L)_{\mu_i}$ the reduction of $\restr{\Delta^t_L}{\bar{J}_L^{-1}(\mu_i)}$, and (3) $(S^t_L){\mu_i}$ the reduction of $\restr{\bar{J}_L}{S_L}^{-1}(\mu_i)$, There is a sequence of reduced simple hybrid time-dependent Lagrangian systems (``Coll.'' stands for collision and
``Red.'' stands for reduction):
\[
\begin{tikzcd}[column sep=.5cm, row
sep=.7cm]
{[\tau_0,\tau_1]} \arrow[d,swap,"\text{coll.}"]\arrow[r,"\text{Red.}"] & (\mathbb{R}\times \T( Q/\mathbb{S}^1),L_{\mu_0},(S^t_L)_{\mu_0},(\Delta^t_L)_{\mu_0}) \arrow[d,"\text{coll.}"]\\
{[\tau_1,\tau_2]} \arrow[d,swap,"\text{coll.}"]\arrow[r,"\text{Red.}"] & (\mathbb{R}\times \T( Q/\mathbb{S}^1),L_{\mu_1},(S^t_L)_{\mu_1},(\Delta^t_L)_{\mu_1}) \arrow[d,"\text{coll.}"]\\
(\dots) \arrow[r,"\text{Red.}"] & (\dots)
\end{tikzcd} 
\]

As in the symplectic case, the fact that the momentum will, in general, change with the collisions makes the reconstruction procedure more challenging. If one wishes, as usual, to use a reduced solution to reconstruct the original dynamics, one needs to compute the reduced hybrid data after each collision. This means that once the reduced solution has been obtained between two collison events, say at $t=\tau_n$ and $t=\tau_{n+1}$, one has to reconstruct this solution to obtain the new momentum after the collision at $\tau_{n+1}$ and use this new momentum to build a new reduced hybrid system whose solution should be obtained until the next collision eventy at $\tau_{n+2}$, and so on (see section \ref{cyclic} for details). 
\end{remark}

{As examples we will consider systems subjected to time-dependent one-sided constraints (see \cite{ibort_geometric_1998}).}
\begin{example}[Billiard with dissipation and moving walls] \label{example_moving}

Consider a particle of mass $m$ in the plane which is free to move inside the surface defined by circle whose
radius varies in time according to a given function $f(t)$, i.e. $x^2+y^2=f(t)$. The surface of the ``billiard'' is assumed to be rough in such a way that the friction is non-linear on the velocities.

The Lagrangian function $L:\R\times \T\mathbb{R}^{2}\to\mathbb{R}$ is given by 
$$\displaystyle{
L(t,x,y,\dot{x},\dot{y})=\frac{m}{2}e^{\frac{ct}{m}}}(\dot x^2+\dot y^2)$$ and $F^L(t,x,y,\dot{x},\dot{y})=F^L_{x}\dd x+F^L_{y}\dd y$ is an external force given by $F^L_{x}=-2ce^{\frac{c}{m}t}(\dot{x}xy-\dot{y}x^2)$, $F^L_{y}=2ce^{\frac{c}{m}t}(\dot{y}xy-\dot{x}y^2)$, for a constant $c>0$. The equations of motion for the particle off the boundary are then 
\begin{equation*}\label{eq:ex11}
c\dot{x}+m\ddot x=-2c(\dot{y}x^2-\dot{x}xy), \quad c\dot{y}+m\ddot y=2c(\dot{x}y^2-\dot{y}xy).
\end{equation*}
The switching surface is the subset of $\mathbb{R}\times \T\mathbb{R}^2\simeq\mathbb{R}\times\mathbb{R}^{2}\times\mathbb{R}^{2}$ given by
\[
S_L^t=(\mathbb{R}\times \T Q)\cap \{x^2+y^2=f(t), (\dot x,\dot y)\cdot (x,y)> \dot{f}(t) \}.
\]
{The first of the equations defining $S_L^t$ means that $(x,y)$ is a point in a circle of radius $f(t)$, while the second means that the component of the velocity pointing ``outwards'' the billiard is positive.}
For simplicity in the definition of the switching surface, we assume that $f(t)$ is increasing: this guarantees the particle only hits the boundary when the boundary is also moving outwards. Under the assumption of an elastic collision, the impact map $(t,x,y,\dot x^{-},\dot y^{-})\mapsto (t,x,y,\dot x^{+},\dot y^{+})$ is given by {(see \cite[Example 10.1]{ibort_geometric_1998})}
\begin{align*}
\dot x^{+}&=\dot x^{-} + \frac{\dot f(t)-2 (x\dot x^{-}+y\dot y^{-})}{f(t)} x,\\
\dot y^{+}&=\dot y^{-} + \frac{\dot f(t)-2 (x\dot x^{-}+y\dot y^{-})}{f(t)} y.
\end{align*}
By introducing polar coordinates $L$ and $F^L$ become
\begin{align*}
L(t,\theta,r,\dot{\theta},\dot{r})&=\frac{m}{2}e^{\frac{c}{m}t}(\dot r^2+ r^2\dot \theta^2),\\ F^L(t,\theta,r,\dot{\theta},\dot{r})&=2e^{\frac{c}{m}t}cr^3\dot{\theta}dr,\end{align*} respectively. $L$ is hyperregular and $L$ and $F^L$ are independent of $\theta$. The forced Euler--Lagrange equations (in polar coordinates) are $$\displaystyle{\ddot{r}=-\frac{2cr^3}{m}\dot{\theta}+r\dot{\theta}^2-\frac{c}{m}\dot{r},\,\,\ddot{\theta}=-\frac{c}{m}\dot{\theta}}.$$

The impact map $\Delta_L^t$, in polar coordinates, takes the form (observe that $2(x\dot x^{-}+y\dot y^{-})$ is nothing but  $2 r \dot r^{-}$):
\begin{align}\label{eq:resetpolar}
(\dot{r}^{+})^{2}&=(\dot{r}^{-})^2
+\frac{r}{f(t)}(\dot{f}(t)-2r\dot{r}^{-})\left(2\dot{r}^{-}+\frac{(\dot{f}(t)-2r\dot{r}^{-})r}{f(t)}\right),\\
\dot{\theta}^{+}&=\dot{\theta}^{-}.\nonumber
\end{align}
Note that the particle bounces on the boundary after the collision, so the ``minus'' square root is considered in $\dot{r}^{+}$ .

Note that the momentum map $(\bar{J}_L)_{F}$ for $\theta$, $$(\bar{J}_L)_{F}(t,r,\dot r,\theta,\dot \theta)= me^{\frac{c}{m}t}r^2\dot\theta$$ is {a hybrid constant of the motion}. Hence, by considering $\mu=me^{\frac{c}{m}t}r^2\dot{\theta}$ (i.e., $\dot{\theta}=\frac{\mu}{mr^2}e^{-\frac{c}{m}t}$) the time-dependent Routhian and the reduced time-dependent force takes the form
\[
R_F^{\mu}(t,r,\dot{r})=\frac{m}{2}e^{\frac{c}{m}t}\dot r^2-\frac{\mu^2}{2mr^2}e^{-\frac{c}{m}t}, \,\, F^L_{\mu}=2cr\frac{\mu} {m}dr\, .
\]
Thus, the time-dependent forced reduced Euler--Lagrange equations for the Routhian $R_F^{\mu}$ are given by
\begin{equation*}
\ddot{r}=\frac{\mu^2}{m^2r^3}e^{-\frac{2c}{m}t}-\frac{2cr\mu}{m^2}e^{-\frac{c}{m}t}-\frac{\dot{r}c}{m}\, .
\end{equation*}
As a matter of fact, $(\bar J_L)_F$ is a hybrid momentum map.
The reduced impact map is given by~\eqref{eq:resetpolar} for $\dot r^+$ (note that the expression drops to the quotient since it only involves $r$, $\dot r$ and $f(t)$). The reduced impact surface is $(S_L)_\mu=\{r^2=f(t), \dot r>0\}$. Hence, we have the simple hybrid time-dependent forced Lagrangian system $\mathscr{L}_{F}=(Q_{\rm red},R_F^{\mu},F_{\mu},(S_{L}^t)_\mu,(\Delta_{L}^t)_\mu)$, with $Q_{\rm red}\simeq \mathbb{R}^+$ parametrized by the coordinate $r$.

\begin{figure}[ht!] 
\centering
	\includegraphics[height=3.9cm]{c0.05r.eps.pdf}\includegraphics[height=3.9cm]{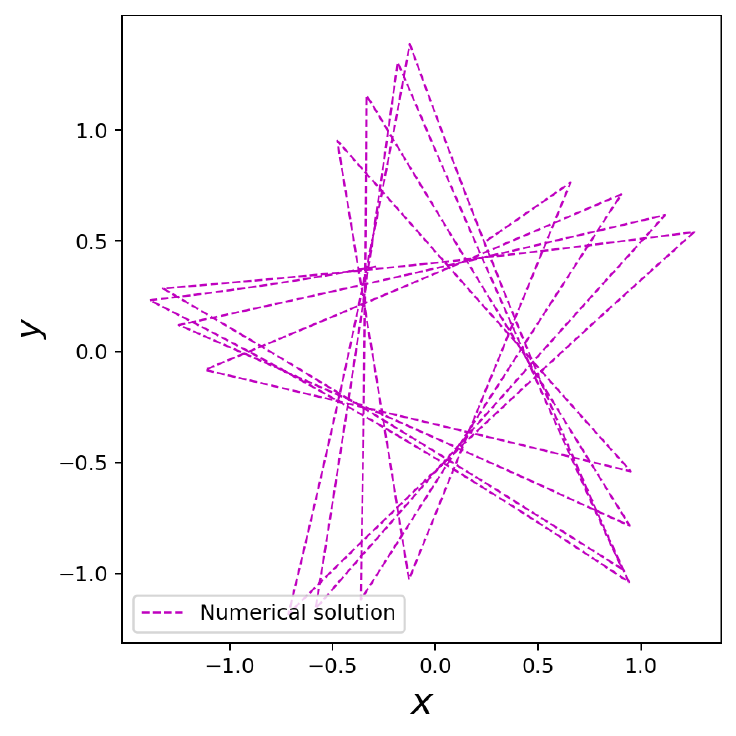}
	\caption{Simulation of Example~\ref{example_moving} for $c=0.005$. The figure in the left corresponds with the reduced trajectory while the figure to the right corresponds with the reconstructed solution}\label{fig:0.005}
\end{figure}
Figures~\ref{fig:0.005} and~\ref{fig:0.10} show numerical results using \textsc{Python} for two different values of the dissipation parameter $c$. The remaining parameters are the same for both simulations: $m=1$, $r(0)=0.5590$, $\dot r(0)=2.8621$, $\theta(0)=1.1071$ (rad), $\dot \theta(0)=-3.0400$ (rad/s), with $f(t)$ given by
\[
f(t)=2-\exp(t/10). 
\]
The reduced dynamics corresponding to $R^\mu_{F}$ is solved numerically (dashed black line) and used to integrate (numerically) the reconstruction equation
\[
\dot\theta=\exp{\left(-\frac{c}{m}t\right)} \frac{\mu} {m r^2}, 
\]
with $\mu$ determined from the initial conditions. 

\begin{figure}[ht!]
\centering
\includegraphics[height=3.9cm]{c01r.eps.pdf}\includegraphics[height=3.9cm]{c01xy.eps.pdf}
 	\caption{Simulation of Example~\ref{example_moving} for $c=0.10$. The figure in the left represents the reduced trajectory while the figure to the right represents the reconstructed solution}\label{fig:0.10}
 \end{figure}\end{example}

\begin{example}[Rolling disk with dissipation hitting a moving wall]
\label{example_disk_time_dependent}

\begin{figure}[h!]
    \centering
    \begin{tikzpicture}    
        \draw[-,blue,ultra thick] (-3,-2)node[left] {\(y=0\)} -- (5,-2) ;
    
        \draw[-,red,ultra thick] (-3,2)node[left] {\(h\leq y=f(t)-R\)} -- (5,2) ;

         \draw[dashed,gray] (-3,1.25)node[left] {\(y=h\)} -- (5,1.25) ;
        
        \draw[->,red,thick] (-0.5,2)--(-0.5,2.5);
        \draw[->,red,thick] (0.5,2)--(0.5,1.5);
    
        \draw[->] (-2.5,-2) -- (3,-2) node[right] {\(x\)};
    
        \draw[->] (-2,-2.5) -- (-2,3) node[above] {\(y\)};
    
        \draw (0,0) circle (1);
    
        \draw[dashed] (0,-1) -- (0,1) node[above] {\(\)};
    
        \draw[-] (0,0) -- ({sin(-40)}, {cos(-40)}) node[midway, above left] {\(\)};
    
        \fill (0,0) circle (1.5pt) node[below left] {\(C\)};
    
        \fill ({sin(-40)}, {cos(-40)}) circle (1.5pt) node[above] {\(P\)};
    
        \draw[thick] (0,0.55) arc[start angle=90, end angle=130, radius=0.55];
    
        \node at (-0.2,0.7) {\(\varphi\)};

        \draw[<->] (0,0)--node[above]{\(R\)}(1,0);
    \end{tikzpicture}
    \caption{Rolling disk with dissipation hitting moving walls from Example~\ref{example_disk_time_dependent}. Here $C$ is the center of the disk, and $P$ a reference point to measure the angle $\varphi$ from the $Oy$ axis.}
    \label{fig:example_disk_moving}
\end{figure}

Consider a homogeneous circular disk of radius $R$ and mass $m$ moving in the vertical plane $xOy$ (see Example \ref{example_disk} and Figure~\ref{fig:example_disk_moving}).

Suppose that there are two rough walls at the axis $y=0$ and at $y=f(t)$, where $f(t)\geq h=2\alpha R$ for some constant $\alpha>1$. When the disk hits one of the walls, the impact map is given by 
\begin{equation}
  \left(\dot x^-, \dot y^-, \dot \varphi^-  \right)
  \mapsto \left(\frac{R^2 \dot x^- +k^2 R \dot\varphi^-}{k^2+R^2}, - \dot y^-, \frac{R \dot x^- +k^2 \dot\varphi^-}{k^2+R^2}  \right),
\end{equation} where
{the switching surface is $S=C_1\cup C_2$, with
\begin{equation}
        \begin{aligned}
                & C_1 = \{(x,y,\varphi,\dot x,\dot y,\dot \varphi)\mid y=R,\,  \dot x =R\dot {\varphi} \hbox{ and } \dot y<0\}\, ,\\
                & C_2 = \{(x,y,\vartheta,p_x,p_y,p_\vartheta)\mid y=f(t)-R,\,  \dot x =R\dot {\varphi} \hbox{ and } \dot y>0\}\, .
        \end{aligned}
\end{equation}
}

 One can check that $(\bar J_L)_F(r,\dot r,\theta,\dot \theta,\varphi,\dot\varphi)= (mr^2\dot\theta,mk^2\dot\varphi)$ is a generalized hybrid momentum map but not a hybrid momentum map, i.e., $(\bar J_L)_F(q_1, \dot q_1^-)=(\bar J_L)_F(q_2, \dot q_2^-)$ implies that $(\bar J_L)_F(q_1, \dot q_1^+)=(\bar J_L)_F(q_2, \dot q_2^+)$  but  $(\bar J_L)_F(q_1, \dot q_1^+)\neq (\bar J_L)_F(q_1, \dot q_1^-)$ .

In polar coordinates, we have 
\begin{equation}
\begin{aligned}
\dot \theta^+
=-\dot \theta^-,
\label{impact_theta_polar_time}
\end{aligned}
\end{equation}
\begin{equation}
\begin{aligned}
   \dot r^+
  & = (2\cos^2 \theta -1) \dot r^- -2r \sin \theta \cos \theta \dot \theta^-,
\end{aligned}
\end{equation}
and
\begin{equation}
\begin{aligned}
\varphi^-
&= \frac{R \left(\cos \varphi\, \dot r^-  -\sin \varphi\, \dot \varphi^- \right) +k^2 \dot\varphi^-}{k^2+R^2}.
\end{aligned}
\end{equation}
{The connected components of the switching surface can be written as
\begin{equation}
\begin{aligned}
        C_1 & = \left\{(r,\theta,\varphi,\dot r,\dot \theta,\dot \varphi)\mid r\sin \theta=R\, , \quad 
        \dot r \cos \theta - r \dot \theta \sin \theta=R\dot {\varphi}
        \quad
        \hbox{and} 
        \quad 
        \dot r \sin \theta + r \dot \theta \cos \theta < 0
        \right\}\, , \\
        C_2 & = \left\{(r,\theta,\varphi,\dot r,\dot \theta,\dot \varphi)\mid r\sin \theta=f(t)-R\, , \quad 
        \dot r \cos \theta - r \dot \theta \sin \theta=R\dot {\varphi}
        \quad
        \hbox{and} 
        \quad 
        \dot r \sin \theta + r \dot \theta \cos \theta > 0
        \right\}\, .
\end{aligned}
\end{equation}
}

Let $(\mu_1^-,\mu_2^-)$ and $(\mu_1^+,\mu_2^+)$ be the value of the momentum map before and after the impact, respectively. We can write $\dot\theta^\pm=\mu_1^\pm/mr^2$ and $\dot \varphi^\pm=\mu_2^\pm/mk^2$, so the reduced switching map is
$$\dot r^- \mapsto  (2\cos^2 \theta -1) \dot r^- -2r \sin \theta \cos \theta \frac{\mu_1}{mr^2},$$
with the relations
$$\mu_1^+=-\mu_1^-,\quad \mu_2^+=\mu_2^-.$$
The reduced switching surface can be written as
\begin{align*}
    S_{(\mu_1,\mu_2)}&=\left\{(r,\dot r)| r\sin \varphi=R \hbox{ or } r\sin \varphi=f(t)-R,
    \right. \\ &\left.\quad
    \hbox{ and } \dot r \cos \varphi - r \frac{\mu_1}{mr^2} \sin \varphi=R\frac{\mu_2}{mk^2},
    \hbox{ for some } \varphi \in [0, 2\pi)\right\}.
\end{align*}

{If the initial value (after the ``$0$-th impact'') of the momentum map is $\mu_0 = (\mu_{0,1}, \mu_{0,2})$, its value after the $j$-th impact will be $\mu_i=\Big((-1)^{j} \mu_{0,1}, \mu_{0,2}\Big)$. Therefore, after the $j$-th impact, the reduced switching map and the connected components of the reduced switching surface will be
\begin{align*}
    \Delta_{\mu_{j}} (\dot r^-) & = (2\cos^2 \theta -1) \dot r^- - (-1)^j\, 2r \sin \theta \cos \theta \frac{\mu_{0,1}}{mr^2}\, ,\\
    C_{1,\mu_{j}}&=\Big\{(r,\dot r)\mid r\sin \gamma=R\, , \
    \dot r \cos \gamma - \frac{(-1)^j\, \mu_{0,1}}{mr} \sin \gamma=R\frac{\mu_{0,2}}{mk^2}\, ,
    \\ &  \quad
    \hbox{ and }
    \dot r \sin \gamma + \frac{(-1)^j\, \mu_{0,1}}{mr} \cos \gamma < 0
    \hbox{ for some } \gamma \in [0, 2\pi)\Big\}\, , \\
    C_{2,\mu_{j}}&=\Big\{(r,\dot r)\mid r\sin \gamma=f(t)-R\, , \
    \dot r \cos \gamma -  \frac{(-1)^j\, \mu_{0,1}}{m} \sin \gamma=R\frac{\mu_{0,2}}{mk^2}\, ,
    \\ &  \quad
    \hbox{ and }
    \dot r \sin \gamma + \frac{(-1)^j\, \mu_{0,1}}{mr} \cos \gamma > 0
    \hbox{ for some } \gamma \in [0, 2\pi)\Big\}\, .
\end{align*}
}

\end{example}

\section{Conclusions and future work} \label{section_conclusions}

The celebrated symplectic reduction of (conservative) mechanical systems with symmetries, due to Marsden, Weinstein and Meyer \cite{abraham_foundations_2008,marsden_reduction_1974,ortega_momentum_2004,meyer_symmetries_1973}, was recently extended for forced autonomous Lagrangian  \cite{de_leon_symmetries_2021} as well as Hamiltonian \cite{de_leon_geometric_2022} systems. In this paper we have gone a step further and considered simple hybrid forced mechanical systems, both autonomous and non-autonomous, with a generalized hybrid momentum map. The main difference in the reduction and reconstruction of these systems with respect to continuous systems, or hybrid systems with a hybrid momentum map, is that, since nonelastic collisions with the switching surface will modify the value of the momentum map (see Examples \ref{example_disk} and \ref{example_disk_time_dependent}, respectively), we have a sequence of reduced simple hybrid forced reduced Hamiltonian systems. In particular, we have considered $\mathbb{S}^1$-invariant hybrid forced autonomous and non-autonomous mechanical systems. 

We plan to extend our results for more general settings. We would like to obtain a reduction procedure for dissipative hybrid systems in the framework of contact geometry. Moreover, we could consider the reduction of systems with both continuous and discrete time dynamics which are not simple hybrid systems. For instance, we could consider $D$ of co-dimension different from 1, or a system having several domains and switching surfaces that separate them \cite{cortes_mechanical_2001,goodman_existence_2020, goodman_variational_2021, chen_new_1997}. 
{An additional intriguing question, which we leave for future research, concerns the long-term behavior after an arbitrary number of impacts have taken place. The symmetries of the system and the impact map may predict periodic motions as in \cite{bloch2017quasivelocities}. We will extend the results in the former paper to the class of hybrid systems studied in this work in a further work.
}

\section*{Acknowledgements}
The authors are grateful to the anonymous referees for their valuable comments. They acknowledge financial support from Grants PID2019-106715GB-C21, PID2022-137909NB-C2 and RED2022-134301-T, funded by MCIN/AEI/ 10.13039/501100011033. Manuel de León and Asier López-Gordón also received support from the Grant CEX2019-000904-S funded by MCIN/AEI/ 10.13039/501100011033. 

\section*{Declaration of interests}

All authors declare that they have no conflicts of interest to disclose.

\section*{Data availability}
The datasets generated during and/or analysed during the current study are available from the corresponding author on reasonable request.






\appendix
\section{Types of symmetries of forced mechanical systems}\label{subsection_symmetries}   

In this Appendix we briefly recall the different types of symmetries a forced Hamiltonian or Lagrangian system can exhibit. See \cite{de_leon_symmetries_2021,lopez-gordon_geometry_2021, lopez-gordon_geometry_2024} for more details.

Let $(L, F^L)$ be a forced Lagrangian system on $\T Q$. A function $f$ on $\T Q$ is called a \myemph{constant of the motion} (or a \myemph{conserved quantity}) for $(L, F^L)$ if it takes a constant value along the trajectories of the system or, in other words, $X_{L,F^L}(f)=0$.

Let $X=X^i \partial/\partial q^i$ be a vector field on $Q$. It has two associated vector fields on $\T Q$, namely its \myemph{vertical lift} $X^v=X^i \partial /\partial \dot q^i$ and its \myemph{complete lift} $X^c=X^i \partial/\partial q^i + \dot q^j \partial X^i/\partial q^j \partial/\partial \dot q^i$ (see \cite{yano_tangent_1973} for an intrinsic definition). 
Then, $X$ is called a 
\begin{itemize}
\item [(i)] \myemph{symmetry of the forced Lagrangian} if $X^c(L)=F^L(X^c)$,
\item [(ii)] \myemph{Lie symmetry} if $[X^c, X_{L,F^L}]=0$,
\item [(iii)] \myemph{Noether symmetry} if $X^c(E_L)+F^L(X^c)=0$ and $\liedv{X^c}\theta_L=\dd f$ for some function $f$ on $\T Q$.
\end{itemize}
Similarly, let $\tilde X$ be a vector field on $\T Q$. Then, $\tilde X$ is called a 
\begin{itemize}
\item [(i)] \myemph{dynamical symmetry} if $[\tilde X, X_{L,F^L}]=0$,
\item [(ii)] \myemph{Noether symmetry} if $\tilde X(E_L)+F^L(\tilde X)=0$ and $\liedv{\tilde X}\theta_L=\dd f$ for some function $f$ on $\T Q$.
\end{itemize}
Moreover, the following relations between symmetries and constants of the motion hold:
\begin{itemize}
\item [(i)] $X$ is a symmetry of the forced Lagrangian if and only if $X^v(L)$ is a constant of the motion.
\item [(ii)] If $X$ satisfies that $\liedv{X^c}\theta_L=\dd f$, then $X$ is a Noether symmetry if and only if $f-X^v(L)$ is a constant of the motion.
\item [(iii)] If $X$ is a Noether symmetry, it is also a Lie symmetry if and only if $\contr{X^c} \dd \beta$=0.
\item [(iv)] $X$ is a Lie symmetry if and only if $X^c$ is a dynamical symmetry.
\item [(v)] $X$ is a Noether symmetry if and only if $X^c$ is a Cartan symmetry.
\item [(vi)] If $\tilde X$ satisfies that $\liedv{\tilde X}\theta_L=\dd f$, then $\tilde X$ is a Cartan symmetry if and only if $f-(\Sendo \tilde X)(L)$ is a constant of the motion. Here $\Sendo$ is the vertical endomorphism.
\item [(vii)] If $\tilde X$ is a Cartan symmetry, it is also a dynamical symmetry if and only if $\contr{\tilde X} \dd \beta=0$.
\end{itemize}

Let now $(H,F)$ be a forced Hamiltonian system on $\cT Q$. A function $f$ on $\cT Q$ is called a \myemph{constant of the motion} (or a \myemph{conserved quantity}) for $(H, F)$ if it takes a constant value along the trajectories of the system or, in other words, $X_{H,F}(f)=0$. A vector field $\hat X$ on $\cT Q$ is called a \myemph{symmetry of the forced Hamiltonian} if $\hat X(H)+F(\hat X)=0$ and $\liedv{X} \theta_Q=\dd f$ for some $f$ on $\cT Q$. If $\hat{X}$ is a symmetry of the forced Hamiltonian, then $f-\theta_Q(\hat X)$ is a constant of the motion. 

In addition, the Hamiltonian and Lagrangian symmetries are related as follows. Suppose that $(L, F^L)$ is the Lagrangian counterpart of $(H, F)$, namely, $H\circ \F L = E_L$ and $\F L^*( F)=F^L$. Let $\tilde X$ be a vector field on $\T Q$ and let $\hat X$ be a $\F L$-related vector field on $\cT Q$, i.e., $\F L_* \circ \tilde X = \hat X\circ \F L$. Then, the following relations hold:
\begin{itemize}
\item [(i)] $[\hat X, X_{H, F}]=0$ if and only if $\tilde X$ is a dynamical symmetry of $(L, F^L)$.
\item [(ii)] $\liedv{\hat X} \theta_Q= \dd f$ if and only if $\liedv{\tilde X} \theta_L = \dd (f \circ \F L)$.
\item [(iii)] If $\liedv{\hat X} \theta_Q= \dd f$, then the following assertions are equivalent:
  \begin{itemize}
    \item [(a)] $\hat X(H)+F(\hat X)=0$,
    \item [(b)] $f-\theta_Q(\hat X)$ is a constant of the motion,
    \item [(c)] $\tilde X(E_L)+F^L(\tilde X)=0$,
    \item [(d)] $f\circ \F L- \theta_L(\tilde X)$ is a constant of the motion.
  \end{itemize}
\end{itemize}

When the set of transformations that leave invariant a forced Lagrangian (or Hamiltonian) system form a Lie group (so the vector fields generating the infinitesimal symmetries close a Lie subalgebra), we can introduce a momentum map, which associates an independent constant of the motion to each of the generators of the Lie algebra (see Section \ref{section_symp_reduction}). This, when the group action ``behaves well'', allows to project the dynamics of the system to a reduced space of less dimensions. 
As a matter of fact, if we know that an (unforced) Hamiltonian system $H$ can be reduced by the action of a Lie group $G$, in order to reduce the forced Hamiltonian system $(H, F)$ it suffices to consider the Lie subgroup $G_F$ whose infinitesimal generators act as symmetries of the forced Hamiltonian (see Remark \ref{remark_symmetries_reduction}).

\let\emph\oldemph
\printbibliography

@book{abraham_foundations_2008,
  title = {Foundations of {{Mechanics}}},
  author = {Abraham, R. and Marsden, Jerrold E.},
  date = {2008},
  publisher = {AMS Chelsea Publishing/American Mathematical Society},
  location = {Providence, RI},
  url = {https://books.google.es/books?id=4Y-ownk6ilsC},
  isbn = {978-0-8218-4438-0},
  lccn = {2008005206}
}

@article{albert_theoreme_1989,
  title = {Le théorème de réduction de Marsden-Weinstein en géométrie cosymplectique et de contact},
  author = {Albert, Claude},
  date = {1989-01-01},
  journaltitle = {J. Geom. Phys.},
  volume = {6},
  number = {4},
  pages = {627--649},
  issn = {0393-0440},
  doi = {10.1016/0393-0440(89)90029-6},
  url = {https://www.sciencedirect.com/science/article/pii/0393044089900296},
  langid = {french}
}

@incollection{ames_geometric_2007,
  title = {On the {{Geometric Reduction}} of {{Controlled Three-Dimensional Bipedal Robotic Walkers}}},
  author = {Ames, Aaron D. and Gregg, Robert D. and Wendel, Eric D. B. and Sastry, Shankar},
  editor = {Bullo, Francesco and Fujimoto, Kenji},
  date = {2007},
  number = {366},
  pages = {183--196},
  publisher = {Springer},
  location = {Berlin},
  url = {https://resolver.caltech.edu/CaltechAUTHORS:20100819-102549602},
  eventtitle = {3rd {{IFAC Workshop}} on {{Lagrangian}} and {{Hamiltonian Methods}} for {{Nonlinear}}},
  isbn = {978-3-540-73889-3},
  issue = {366}
}

@inproceedings{ames_hybrid_2006,
  title = {Hybrid Cotangent Bundle Reduction of Simple Hybrid Mechanical Systems with Symmetry},
  booktitle = {2006 {{American Control Conference}}},
  author = {Ames, A.D. and Sastry, S.},
  date = {2006},
  pages = {6 pp.},
  publisher = {IEEE},
  location = {Minneapolis, MN, USA},
  doi = {10.1109/ACC.2006.1656622},
  url = {http://ieeexplore.ieee.org/document/1656622/},
  eventtitle = {2006 {{American Control Conference}}},
  isbn = {978-1-4244-0209-0},
  langid = {english}
}

@inproceedings{ames_hybrid_2006-1,
  title = {Hybrid {{Routhian}} Reduction of {{Lagrangian}} Hybrid Systems},
  booktitle = {2006 {{American Control Conference}}},
  author = {Ames, A.D. and Sastry, S.},
  date = {2006-06},
  pages = {6 pp.},
  publisher = {IEEE},
  location = {Minneapolis, MN, USA},
  issn = {2378-5861},
  doi = {10.1109/ACC.2006.1656621},
  eventtitle = {2006 {{American Control Conference}}}
}

@book{brogliato_nonsmooth_1996,
  title = {Nonsmooth {{Impact Mechanics}}},
  author = {Brogliato, Bernard},
  date = {1996},
  series = {Lecture {{Notes}} in {{Control}} and {{Information Sciences}}},
  volume = {220},
  publisher = {Springer-Verlag},
  location = {London},
  doi = {10.1007/BFb0027733},
  url = {http://link.springer.com/10.1007/BFb0027733},
  isbn = {978-3-540-76079-5},
  langid = {english}
}

@article{cantrijn_gradient_1992,
  title = {Gradient Vector Fields on Cosymplectic Manifolds},
  author = {Cantrijn, F and family=León, given=Manuel, prefix=de, useprefix=false and Lacomba, E A},
  date = {1992-01-07},
  journaltitle = {J. Phys. A: Math. Gen.},
  volume = {25},
  number = {1},
  pages = {175--188},
  issn = {0305-4470, 1361-6447},
  doi = {10.1088/0305-4470/25/1/022},
  url = {https://iopscience.iop.org/article/10.1088/0305-4470/25/1/022},
  langid = {english}
}

@article{chen_new_1997,
  title = {A New Classification of Non-Holonomic Constraints},
  author = {Chen, B. and Wang, L.-S. and Chu, S.-S. and Chou, W.-T.},
  date = {1997-03-08},
  journaltitle = {Proc. R. Soc. Lond. A},
  volume = {453},
  number = {1958},
  pages = {631--642},
  publisher = {Royal Society},
  doi = {10.1098/rspa.1997.0035},
  url = {https://royalsocietypublishing.org/doi/10.1098/rspa.1997.0035}
}

@inproceedings{clark_bouncing_2019,
  title = {The {{Bouncing Penny}} and {{Nonholonomic Impacts}}},
  booktitle = {2019 {{IEEE}} 58th {{Conference}} on {{Decision}} and {{Control}} ({{CDC}})},
  author = {Clark, William and Bloch, Anthony},
  date = {2019-12},
  pages = {2114--2119},
  location = {Nice, France},
  issn = {2576-2370},
  doi = {10.1109/CDC40024.2019.9029545},
  eventtitle = {2019 {{IEEE}} 58th {{Conference}} on {{Decision}} and {{Control}} ({{CDC}})}
}

@article{colombo_note_2020,
  title = {A Note on {{Hybrid Routh}} Reduction for Time-Dependent {{Lagrangian}} Systems},
  author = {Colombo, Leonardo J. and Irazú, María Emma Eyrea and García-Toraño Andrés, Eduardo},
  date = {2020},
  journaltitle = {J. Geom. Mech.},
  volume = {12},
  number = {2},
  pages = {309},
  doi = {10.3934/jgm.2020014},
  url = {https://www.aimsciences.org/article/doi/10.3934/jgm.2020014},
  langid = {english}
}

@article{colombo_symmetries_2020,
  title = {Symmetries and Periodic Orbits in Simple Hybrid {{Routhian}} Systems},
  author = {Colombo, Leonardo J. and Eyrea Irazú, María Emma},
  date = {2020-05-01},
  journaltitle = {Nonlinear Analysis: Hybrid Systems},
  volume = {36},
  pages = {100857},
  issn = {1751-570X},
  doi = {10.1016/j.nahs.2020.100857},
  url = {https://www.sciencedirect.com/science/article/pii/S1751570X20300042},
  langid = {english}
}

@inproceedings{colombo_time_2018,
  title = {Time Reversal Symmetries and Zero Dynamics for Simple Hybrid {{Hamiltonian}} Control Systems},
  booktitle = {2018 {{Annual American Control Conference}} ({{ACC}})},
  author = {Colombo, Leonardo and Clark, William and Bloch, Anthony},
  date = {2018-06},
  pages = {2218--2223},
  location = {Milwaukee, WI, USA},
  issn = {2378-5861},
  doi = {10.23919/ACC.2018.8431672},
  eventtitle = {2018 {{Annual American Control Conference}} ({{ACC}})}
}

@article{cortes_hamiltonian_2006,
  title = {Hamiltonian Theory of Constrained Impulsive Motion},
  author = {Cortés, Jorge and Vinogradov, Alexandre M.},
  date = {2006-04-01},
  journaltitle = {J. Math. Phys.},
  volume = {47},
  number = {4},
  pages = {042905},
  publisher = {American Institute of Physics},
  issn = {0022-2488},
  doi = {10.1063/1.2192974},
  url = {https://aip.scitation.org/doi/10.1063/1.2192974}
}

@article{cortes_mechanical_2001,
  title = {Mechanical Systems Subjected to Generalized Non-Holonomic Constraints},
  author = {Cortés, J. and family=León, given=Manuel, prefix=de, useprefix=true and Martín de Diego, David and Martínez, S.},
  date = {2001-03-08},
  journaltitle = {Proc. R. Soc. Lond. A},
  volume = {457},
  number = {2007},
  pages = {651--670},
  issn = {1364-5021, 1471-2946},
  doi = {10.1098/rspa.2000.0686},
  url = {https://royalsocietypublishing.org/doi/10.1098/rspa.2000.0686},
  langid = {english}
}

@article{crampin_rouths_2008,
  title = {Routh’s Procedure for Non-{{Abelian}} Symmetry Groups},
  author = {Crampin, M. and Mestdag, T.},
  date = {2008-03-01},
  journaltitle = {J. Math. Phys.},
  volume = {49},
  number = {3},
  pages = {032901},
  publisher = {American Institute of Physics},
  issn = {0022-2488},
  doi = {10.1063/1.2885077},
  url = {https://aip.scitation.org/doi/10.1063/1.2885077}
}

@article{de_leon_cosymplectic_1993,
  title = {Cosymplectic Reduction for Singular Momentum Maps},
  author = {family=León, given=Manuel, prefix=de, useprefix=true and Saralegi, M.},
  date = {1993-10},
  journaltitle = {J. Phys. A: Math. Gen.},
  volume = {26},
  number = {19},
  pages = {5033--5043},
  publisher = {IOP Publishing},
  issn = {0305-4470},
  doi = {10.1088/0305-4470/26/19/032},
  url = {https://doi.org/10.1088/0305-4470/26/19/032},
  langid = {english}
}

@article{de_leon_geometric_2022,
  title = {Geometric {{Hamilton}}–{{Jacobi}} Theory for Systems with External Forces},
  author = {family=León, given=Manuel, prefix=de, useprefix=true and Lainz, Manuel and López-Gordón, Asier},
  date = {2022-02-01},
  journaltitle = {J. Math. Phys.},
  volume = {63},
  number = {2},
  pages = {022901},
  publisher = {American Institute of Physics},
  issn = {0022-2488},
  doi = {10.1063/5.0073214},
  url = {https://aip.scitation.org/doi/10.1063/5.0073214}
}

@article{de_leon_symmetries_2021,
  title = {Symmetries, Constants of the Motion, and Reduction of Mechanical Systems with External Forces},
  author = {family=León, given=Manuel, prefix=de, useprefix=true and Lainz, Manuel and López-Gordón, Asier},
  date = {2021-04-01},
  journaltitle = {J. Math. Phys.},
  volume = {62},
  number = {4},
  pages = {042901},
  publisher = {American Institute of Physics},
  issn = {0022-2488},
  doi = {10.1063/5.0045073},
  url = {https://aip.scitation.org/doi/10.1063/5.0045073}
}

@inproceedings{eyrea_irazu_hybrid_2022,
  title = {Hybrid {{Routhian}} Reduction for Simple Hybrid Forced {{Lagrangian}} Systems},
  booktitle = {2022 {{European Control Conference}} ({{ECC}})},
  author = {Eyrea Irazú, María Emma and López-Gordón, Asier and family=León, given=Manuel, prefix=de, useprefix=true and Colombo, Leonardo J.},
  date = {2022-07},
  pages = {345--350},
  doi = {10.23919/ECC55457.2022.9838077},
  eventtitle = {2022 {{European Control Conference}} ({{ECC}})}
}

@article{eyrea_irazu_reduction_2021,
  title = {Reduction by {{Symmetries}} of {{Simple Hybrid Mechanical Systems}}},
  shorttitle = {Reduction by {{Symmetries}} of {{Simple Hybrid Mechanical Systems}}⁎⁎{{The}} Project That Gave Rise to These Results Received the Support of a Fellowship from “La {{Caixa}}” {{Foundation}} ({{ID}} 100010434). {{The}} Fellowship Code Is {{LCF}}/{{BQ}}/{{PI19}}/11690016. {{L}}. {{Colombo}} Was Partially Supported by {{Ministe-rio}} de {{Economia}}, {{Industria}} y {{Competitividad}} ({{MINEICO}}, {{Spain}}) under Grant {{MTM2016-76702-P}}; ”{{Severo Ochoa Programme}} for {{Centres}} of {{Excellence}}” in {{R}}\&{{D}} ({{SEV-2015-0554}}) and by {{I-Link Project}} ({{Ref}}},
  author = {Eyrea Irazú, María Emma and Colombo, Leonardo and Bloch, Anthony},
  date = {2021-01-01},
  journaltitle = {7th IFAC Workshop on Lagrangian and Hamiltonian Methods for Nonlinear Control LHMNC 2021},
  series = {{{IFAC-PapersOnLine}}},
  volume = {54},
  number = {19},
  pages = {94--99},
  doi = {10.1016/j.ifacol.2021.11.061},
  url = {https://www.sciencedirect.com/science/article/pii/S240589632102084X},
  langid = {english}
}

@article{garcia-torano_andres_aspects_2014,
  title = {Aspects of Reduction and Transformation of {{Lagrangian}} Systems with Symmetry},
  author = {García-Toraño Andrés, Eduardo and Langerock, Bavo and Cantrijn, Frans},
  date = {2014},
  journaltitle = {J. Geom. Mech.},
  volume = {6},
  number = {1},
  pages = {1},
  doi = {10.3934/jgm.2014.6.1},
  url = {https://www.aimsciences.org/article/doi/10.3934/jgm.2014.6.1},
  langid = {english}
}

@book{godbillon_geometrie_1969,
  title = {Géométrie Différentielle et Mécanique Analytique},
  author = {Godbillon, C.},
  date = {1969},
  series = {Collection {{Méthodes}}},
  publisher = {Hermann},
  location = {Paris},
  url = {https://books.google.es/books?id=0VrvAAAAMAAJ},
  lccn = {lc76420748}
}

@book{goebel_hybrid_2012,
  title = {Hybrid {{Dynamical Systems}}},
  author = {Goebel, Rafal and Sanfelice, Ricardo G.},
  date = {2012-03-18},
  publisher = {Princeton University Press},
  location = {Princeton, NJ},
  url = {https://press.princeton.edu/books/hardcover/9780691153896/hybrid-dynamical-systems},
  isbn = {978-0-691-15389-6},
  langid = {english},
  pagetotal = {232}
}

@book{goldstein_classical_1980,
  title = {Classical Mechanics},
  author = {Goldstein, Herbert},
  date = {1980},
  series = {Addison-{{Wesley Series}} in {{Physics}}},
  edition = {Second edition},
  publisher = {Addison-Wesley Publishing Co., Reading, Mass.},
  location = {Reading, MA},
  url = {https://mathscinet.ams.org/mathscinet-getitem?mr=575343},
  isbn = {978-0-201-02918-5},
  mrnumber = {575343},
  pagetotal = {xiv+672}
}

@article{goodman_existence_2020,
  title = {On the {{Existence}} and {{Uniqueness}} of {{Poincaré Maps}} for {{Systems With Impulse Effects}}},
  author = {Goodman, Jacob R. and Colombo, Leonardo Jesus},
  date = {2020-04},
  journaltitle = {IEEE Transactions on Automatic Control},
  volume = {65},
  number = {4},
  pages = {1815--1821},
  issn = {1558-2523},
  doi = {10.1109/TAC.2019.2941446},
  eventtitle = {{{IEEE Transactions}} on {{Automatic Control}}}
}

@article{goodman_variational_2021,
  title = {Variational {{Obstacle Avoidance}} with {{Applications}} to {{Interpolation Problems}} in {{Hybrid Systems}}},
  author = {Goodman, Jacob and Colombo, Leonardo},
  date = {2021-01-01},
  journaltitle = {IFAC-PapersOnLine},
  series = {7th {{IFAC Workshop}} on {{Lagrangian}} and {{Hamiltonian Methods}} for {{Nonlinear Control LHMNC}} 2021},
  volume = {54},
  number = {19},
  pages = {82--87},
  issn = {2405-8963},
  doi = {10.1016/j.ifacol.2021.11.059},
  url = {https://www.sciencedirect.com/science/article/pii/S2405896321020826},
  langid = {english}
}

@article{grabowska_geometry_2019,
  title = {Geometry of {{Routh}} Reduction},
  author = {Grabowska, Katarzyna and Urbański, Pawel},
  date = {2019},
  journaltitle = {J. Geom. Mech.},
  volume = {11},
  number = {1},
  pages = {23--44},
  issn = {1941-4897},
  doi = {10.3934/jgm.2019002},
  url = {http://aimsciences.org//article/doi/10.3934/jgm.2019002},
  langid = {english}
}

@article{holmes_dynamics_2006,
  title = {The {{Dynamics}} of {{Legged Locomotion}}: {{Models}}, {{Analyses}}, and {{Challenges}}},
  shorttitle = {The {{Dynamics}} of {{Legged Locomotion}}},
  author = {Holmes, Philip and Full, Robert J. and Koditschek, Dan and Guckenheimer, John},
  date = {2006-01},
  journaltitle = {SIAM Rev.},
  volume = {48},
  number = {2},
  pages = {207--304},
  publisher = {{Society for Industrial and Applied Mathematics}},
  issn = {0036-1445},
  doi = {10.1137/S0036144504445133},
  url = {https://epubs.siam.org/doi/10.1137/S0036144504445133}
}

@article{ibort_geometric_1998,
  title = {Geometric Formulation of Mechanical Systems Subjected to Time-Dependent One-Sided Constraints},
  author = {Ibort, Alberto and family=León, given=Manuel, prefix=de, useprefix=true and Lacomba, Ernesto A and Marrero, Juan C and Martín de Diego, David and Pitanga, Paulo},
  date = {1998-03-20},
  journaltitle = {J. Phys. A: Math. Gen.},
  volume = {31},
  number = {11},
  pages = {2655--2674},
  issn = {0305-4470, 1361-6447},
  doi = {10.1088/0305-4470/31/11/014},
  url = {https://iopscience.iop.org/article/10.1088/0305-4470/31/11/014}
}

@article{ibort_geometric_2001,
  title = {Geometric Formulation of {{Carnot}}'s Theorem},
  author = {Ibort, Alberto and family=León, given=Manuel, prefix=de, useprefix=true and Lacomba, Ernesto A and Marrero, Juan C and Martín de Diego, David and Pitanga, Paulo},
  date = {2001-03-02},
  journaltitle = {J. Phys. A: Math. Gen.},
  volume = {34},
  number = {8},
  pages = {1691--1712},
  issn = {0305-4470, 1361-6447},
  doi = {10.1088/0305-4470/34/8/314},
  url = {https://iopscience.iop.org/article/10.1088/0305-4470/34/8/314}
}

@article{ibort_mechanical_1997,
  title = {Mechanical Systems Subjected to Impulsive Constraints},
  author = {Ibort, Alberto and family=León, given=Manuel, prefix=de, useprefix=true and Lacomba, Ernesto A and Martín de Diego, David and Pitanga, Paulo},
  date = {1997-08-21},
  journaltitle = {J. Phys. A: Math. Gen.},
  volume = {30},
  number = {16},
  pages = {5835--5854},
  issn = {0305-4470, 1361-6447},
  doi = {10.1088/0305-4470/30/16/024},
  url = {https://iopscience.iop.org/article/10.1088/0305-4470/30/16/024}
}

@article{johnson_simple_1994,
  title = {Simple Hybrid Systems},
  author = {Johnson, Stewart D.},
  date = {1994-12-01},
  journaltitle = {Int. J. Bifurcation Chaos},
  volume = {04},
  number = {06},
  pages = {1655--1665},
  publisher = {World Scientific Publishing Co.},
  issn = {0218-1274},
  doi = {10.1142/S021812749400126X},
  url = {https://www.worldscientific.com/doi/10.1142/S021812749400126X}
}

@article{langerock_routh_2011,
  title = {Routh {{Reduction}} by {{Stages}}},
  author = {Langerock, Bavo and Mestdag, Tom and Vankerschaver, Joris},
  date = {2011-11-29},
  journaltitle = {SIGMA. Symmetry, Integrability and Geometry: Methods and Applications},
  volume = {7},
  pages = {109},
  publisher = {{SIGMA. Symmetry, Integrability and Geometry: Methods and Applications}},
  issn = {18150659},
  doi = {10.3842/SIGMA.2011.109},
  url = {http://www.emis.de/journals/SIGMA/2011/109/},
  langid = {english}
}

@article{langerock_routh_2012,
  title = {Routh Reduction and the Class of Magnetic {{Lagrangian}} Systems},
  author = {Langerock, B. and García-Toraño Andrés, Eduardo and Cantrijn, F.},
  date = {2012-06-01},
  journaltitle = {J. Math. Phys.},
  volume = {53},
  number = {6},
  pages = {062902},
  publisher = {American Institute of Physics},
  issn = {0022-2488},
  doi = {10.1063/1.4723841},
  url = {https://aip.scitation.org/doi/10.1063/1.4723841}
}

@article{langerock_routhian_2010,
  title = {Routhian Reduction for Quasi-Invariant {{Lagrangians}}},
  author = {Langerock, B. and Cantrijn, F. and Vankerschaver, J.},
  date = {2010-02-01},
  journaltitle = {J. Math. Phys.},
  volume = {51},
  number = {2},
  pages = {022902},
  publisher = {American Institute of Physics},
  issn = {0022-2488},
  doi = {10.1063/1.3277181},
  url = {https://aip.scitation.org/doi/10.1063/1.3277181}
}

@inproceedings{lee_geometric_2013,
  title = {Geometric Control of Cooperating Multiple Quadrotor {{UAVs}} with a Suspended Payload},
  booktitle = {52nd {{IEEE Conference}} on {{Decision}} and {{Control}}},
  author = {Lee, Taeyoung and Sreenath, Koushil and Kumar, Vijay},
  date = {2013-12},
  pages = {5510--5515},
  publisher = {IEEE},
  location = {Firenze},
  doi = {10.1109/CDC.2013.6760757},
  url = {http://ieeexplore.ieee.org/document/6760757/},
  eventtitle = {2013 {{IEEE}} 52nd {{Annual Conference}} on {{Decision}} and {{Control}} ({{CDC}})},
  isbn = {978-1-4673-5717-3 978-1-4673-5714-2 978-1-4799-1381-7},
  langid = {english}
}

@incollection{lee_quotient_2012,
  title = {Quotient {{Manifolds}}},
  booktitle = {Introduction to {{Smooth Manifolds}}},
  author = {Lee, John M.},
  editor = {Lee, John M.},
  date = {2012},
  series = {Graduate {{Texts}} in {{Mathematics}}},
  pages = {540--563},
  publisher = {Springer},
  location = {New York, NY},
  doi = {10.1007/978-1-4419-9982-5_21},
  url = {https://doi.org/10.1007/978-1-4419-9982-5_21},
  isbn = {978-1-4419-9982-5},
  langid = {english}
}

@thesis{lopez-gordon_geometry_2021,
  type = {MSc Thesis},
  title = {The Geometry of {{Rayleigh}} Dissipation},
  author = {López-Gordón, Asier},
  date = {2021-07-08},
  eprint = {2107.03780},
  eprinttype = {arXiv},
  institution = {Universidad Autónoma de Madrid},
  url = {http://arxiv.org/abs/2107.03780}
}

@thesis{lopez-gordon_geometry_2024,
  type = {phdthesis},
  title = {The Geometry of Dissipation},
  author = {López-Gordón, Asier},
  date = {2024-09-18},
  eprint = {2409.11947},
  eprinttype = {arXiv},
  institution = {Universidad Autónoma de Madrid},
  url = {http://arxiv.org/abs/2409.11947}
}

@book{marsden_hamiltonian_2007,
  title = {Hamiltonian {{Reduction}} by {{Stages}}},
  author = {Marsden, J. E. and Misiolek, Gerard and Ortega, Juan-Pablo and Perlmutter, Matthew and Ratiu, Tudor},
  date = {2007},
  series = {Lecture {{Notes}} in {{Mathematics}}},
  publisher = {Springer-Verlag},
  location = {Berlin; Heidelberg},
  doi = {10.1007/978-3-540-72470-4},
  url = {https://www.springer.com/gp/book/9783540724698},
  isbn = {978-3-540-72469-8},
  langid = {english}
}

@book{marsden_introduction_1999,
  title = {Introduction to {{Mechanics}} and {{Symmetry}}: {{A Basic Exposition}} of {{Classical Mechanical Systems}}},
  shorttitle = {Introduction to {{Mechanics}} and {{Symmetry}}},
  author = {Marsden, Jerrold E. and Ratiu, Tudor S.},
  editor = {Marsden, Jerrold E. and Sirovich, L. and Golubitsky, M. and Jäger, W.},
  editortype = {redactor},
  date = {1999},
  series = {Texts in {{Applied Mathematics}}},
  volume = {17},
  publisher = {Springer New York},
  location = {New York, NY},
  doi = {10.1007/978-0-387-21792-5},
  url = {http://link.springer.com/10.1007/978-0-387-21792-5},
  isbn = {978-1-4419-3143-6 978-0-387-21792-5},
  langid = {english}
}

@book{marsden_lectures_1992,
  title = {Lectures on {{Mechanics}}},
  author = {Marsden, Jerrold E.},
  date = {1992},
  series = {London {{Mathematical Society Lecture Note Series}}},
  publisher = {Cambridge University Press},
  location = {Cambridge},
  doi = {10.1017/CBO9780511624001},
  url = {https://www.cambridge.org/core/books/lectures-on-mechanics/49B6A89B9110CAA42E4C169C1A81625D},
  isbn = {978-0-521-42844-6}
}

@article{marsden_reduction_1974,
  title = {Reduction of Symplectic Manifolds with Symmetry},
  author = {Marsden, Jerrold E. and Weinstein, Alan J.},
  date = {1974-02},
  journaltitle = {Rep. Math. Phys.},
  volume = {5},
  number = {1},
  pages = {121--130},
  publisher = {Elsevier},
  issn = {0034-4877},
  url = {https://resolver.caltech.edu/CaltechAUTHORS:20100910-101924768},
  issue = {1},
  langid = {english}
}

@book{haddad2006impulsive,
  title={Impulsive and hybrid dynamical systems: stability, dissipativity, and control},
  author={Haddad, Wassim M and Chellaboina, VijaySekhar and Nersesov, Sergey G},
  year={2006},
  publisher={Princeton University Press}
}

@inproceedings{bloch2017quasivelocities,
  title={Quasivelocities and symmetries in simple hybrid systems},
  author={Bloch, Anthony and Clark, William and Colombo, Leonardo},
  booktitle={2017 IEEE 56th Annual Conference on Decision and Control (CDC)},
  pages={1529--1534},
  year={2017},
  organization={IEEE}
}

@article{william2020poincare,
  title={A Poincar{\'e}-Bendixson theorem for hybrid systems},
  author={Clark, and Bloch, Anthony and Colombo, Leonardo},
  journal={Mathematical Control and Related Fields},
  volume={10},
  number={1},
  pages={27--45},
  year={2020},
  publisher={Mathematical Control and Related Fields}
}

@incollection{meyer_symmetries_1973,
  title = {Symmetries and Integrals in Mechanics},
  booktitle = {Dynamical Systems ({{Proc}}. {{Sympos}}., {{Univ}}. {{Bahia}}, {{Salvador}}, 1971)},
  author = {Meyer, Kenneth R.},
  date = {1973},
  pages = {259--272},
  publisher = {Academic Press, New York-London},
  mrnumber = {331427}
}

@book{ortega_momentum_2004,
  title = {Momentum {{Maps}} and {{Hamiltonian Reduction}}},
  author = {Ortega, Juan-Pablo and Ratiu, Tudor S.},
  date = {2004},
  publisher = {Birkhäuser Boston},
  location = {Boston, MA},
  doi = {10.1007/978-1-4757-3811-7},
  url = {http://link.springer.com/10.1007/978-1-4757-3811-7},
  isbn = {978-1-4757-3813-1 978-1-4757-3811-7},
  langid = {english}
}

@book{pars_treatise_1965,
  title = {A Treatise on Analytical Dynamics},
  author = {Pars, L. A.},
  date = {1965},
  publisher = {Heinemann Educational Books Ltd,},
  location = {London},
  mrnumber = {208873},
  pagetotal = {xxi+641}
}

@book{van_der_schaft_introduction_2000,
  title = {An Introduction to Hybrid Dynamical Systems},
  author = {family=Schaft, given=Arjan, prefix=van der, useprefix=true and {Schumacher, Hans}},
  date = {2000},
  journaltitle = {Lecture Notes in Control and Information Sciences},
  volume = {251},
  publisher = {Springer},
  location = {London},
  issn = {1610-7411},
  doi = {10.1007/bfb0109998},
  url = {https://research.utwente.nl/en/publications/an-introduction-to-hybrid-dynamical-systems(e4de97e1-c133-4dc9-9ffa-1394e35ae6d9).html},
  langid = {english}
}

@book{westervelt_feedback_2018,
  title = {Feedback {{Control}} of {{Dynamic Bipedal Robot Locomotion}}},
  author = {Westervelt, Eric R. and Grizzle, Jessy W. and Chevallereau, Christine and Choi, Jun Ho and Morris, Benjamin},
  date = {2018-10-31},
  publisher = {CRC Press},
  location = {Boca Raton},
  doi = {10.1201/9781420053739},
  isbn = {978-1-315-21942-4},
  pagetotal = {528}
}

@book{yano_tangent_1973,
  title = {Tangent and Cotangent Bundles ; Differential Geometry},
  author = {Yano, Kentaro and Ishihara, Shigeru},
  date = {1973},
  publisher = {Marcel Dekker, Inc.},
  location = {New York, NY},
  langid = {english}
}

\end{document}